\documentclass[letterpaper,12pt]{article}

\usepackage{lmodern}
\usepackage[margin=1in]{geometry}
\usepackage[authoryear,round]{natbib}
\usepackage{graphicx}
\usepackage{amsmath,amssymb}
\usepackage[colorlinks=true,linkcolor=blue,citecolor=blue,urlcolor=blue]{hyperref}
\usepackage{booktabs}
\usepackage{xspace}
\usepackage{etoolbox}
\usepackage{xcolor}
\usepackage{fancyhdr}

\fancypagestyle{plain}{\fancyhf{}\fancyfoot[C]{\thepage}\fancyfoot[R]{\scriptsize SAND2026-21959O}}

\def\tsc#1{\csdef{#1}{\textsc{\lowercase{#1}}\xspace}}
\tsc{WGM}
\tsc{QE}
\begin{document}

\begin{center}
\vspace*{1em}
\fontsize{24.88}{30}\fontseries{bx}\selectfont The Role of Source Geometry and Atmospheric Propagation in Global Bolide Infrasound Detectability\par\normalfont
\vskip 1.5em
{\large M.\ Ronac Giannone\textsuperscript{1,*} \hspace{1.5em} E.\ A.\ Silber\textsuperscript{1}\par}
\vskip 1em
{\small\textsuperscript{1}Sandia National Laboratories, 1515 Eubank SE, Albuquerque, New Mexico 87123, United States\\[6pt]
\textsuperscript{*}Corresponding author: \texttt{miroronac24@gmail.com}}
\end{center}
\vskip 1.5em

\begin{abstract}
Global infrasound monitoring provides a persistent means of detecting energetic bolide atmospheric entries, complementing optical observations and extending coverage over remote regions. We present a global assessment of the physical factors governing bolide infrasound detectability by correlating 623 bolide events reported by the Center for Near-Earth Object Studies between 2007 and 2025 with waveform data from the International Monitoring System. We identify 311 events with confirmed infrasound detections, corresponding to a detection rate of approximately 50\%, substantially higher than inferred from earlier surveys, reflecting both the maturation of the global infrasound network and advances in automated, multi-frequency array processing. Analysis of flight parameters shows that infrasound detectability is selective rather than uniform across the bolide population. Detected events are preferentially associated with steeper entry angles and lower-altitude energy deposition, while shallow, high-altitude trajectories are less consistently observed. Very high-energy events remain detectable regardless of geometry, but for the more common lower-energy regime, observability depends on specific combinations of entry parameters and propagation conditions. This geometric dependence persists across comparable energy ranges and atmospheric conditions, indicating that entry angle exerts a primary control on detectability, with energy and propagation acting as secondary modulating factors. These results provide new physical constraints on bolide-atmosphere interactions and improve interpretation of global infrasound observations for planetary defense and atmospheric-entry studies.
\end{abstract}

\vfill
\begin{center}
{\large\textbf{\textcolor{blue}{Accepted for publication in \textit{Icarus} on 27 May 2026, Special issue: Meteoroids 2025 -- Recent Advances in Meteor Science\\
DOI: \href{https://doi.org/10.1016/j.icarus.2026.117194}{10.1016/j.icarus.2026.117194}}}}
\end{center}

\newpage

\section{Introduction}\label{introduction}

Energetic atmospheric entry of meteoroids and small asteroids produces some of the most extreme natural shock-generating phenomena in Earth's atmosphere. As these bodies penetrate progressively denser atmospheric layers at hypersonic velocities, their kinetic energy is transferred to the surrounding gas through a combination of complex physical and chemical processes, including sputtering, ablation, vaporization, ionization, and nonequilibrium gas-phase chemistry \citep{Bronshten1983, Ceplecha1998, Silber2018}. These processes generate strong shock waves in the surrounding flow field, accompanied by intense luminous emissions that trace the meteoroid's passage through the atmosphere \citep{Silber2018}. Exceptionally bright meteors produced during this process are classified as fireballs, conventionally defined as events brighter than Venus (visual magnitude --4). The term \textit{bolide} is commonly applied to highly energetic fireballs, particularly those involving explosive disruption, although the terminology is often used interchangeably in the literature \citep{Ceplecha1998}. In this work, we use \textit{bolide} to refer to energetic meteoroid entry events that deposit substantial acoustic energy into the atmosphere.

The low-frequency component of this acoustic radiation, known as infrasound, propagates at frequencies below 20 Hz and possesses unique diagnostic utility due to its ability to travel over global distances with minimal attenuation \citep{Evans1972, Sutherland2004}. As a result, bolide-generated infrasound can propagate hundreds to thousands of kilometers when atmospheric conditions are favorable \citep{ReVelle1997, Drob2003, Marcillo2013}. Unlike optical observations, which are constrained by cloud cover, daylight conditions, limited fields of view, and line-of-sight geometry, infrasound sensors operate continuously and omnidirectionally, providing over-the-horizon detection capability and a persistent global watch for atmospheric energy deposition. This makes infrasound a critical complementary technology to satellite and ground-based optical systems, filling observational gaps over open oceans and remote regions.

The acoustic waveform carries a physical imprint of the source mechanism, allowing infrasound to serve as a robust proxy for bolide energetics and fragmentation history in cases where optical or satellite data are unavailable or obscured \citep{Silber2009, Silber2019, Clemente2025}. Infrasound has also proven valuable for characterizing a wide range of energetic natural and anthropogenic sources when integrated with complementary observations such as seismic data. Joint seismoacoustic analyses have improved constraints on various source parameters for near-surface events including volcanic eruptions, earthquakes, as well as chemical and nuclear explosions \citep{Assink2018, Matoza2018, Koch2019, Pasyanos2019, Ronac2024, Park2025}. These multi-sensor approaches highlight the broader potential of infrasound to improve event interpretation beyond what is possible from any single data modality alone.

Detecting and characterizing these events, however, presents unique challenges compared to stationary infrasound sources such as point-source explosions or volcanic eruptions. Bolides are hypersonic moving sources that generate an extremely narrow Mach cone, often approximated as a cylindrical wavefront, rather than a spherical point-source shock \citep{Silber2018}. This distinct source geometry produces a highly directional radiation pattern that evolves rapidly along the trajectory \citep{Brown2013, Popova2013, Wilson2025}. Understanding these moving-source dynamics is critical not only for natural impactors but also for artificial re-entries, such as space debris and controlled spacecraft returns, which share similar hypersonic flight and shock-generation characteristics \citep{ReVelle2007, Yamamoto2011, Ishihara2012, Nishikawa2022, Clemente2025, Silber2024b, Silber2025a, Hatty2026}.

Despite decades of observation, the conditions under which energetic meteoroid entry produces infrasound detectable at regional to global distances remain insufficiently constrained. Atmospheric energy deposition is highly variable and depends on entry velocity, trajectory geometry, and the altitude and vertical extent of energy release \citep{Pilger2015, Silber2025b, Silber2025c, Scamfer2026}. Subsequent propagation to infrasound sensors, and therefore overall detectability, is also strongly influenced by atmospheric conditions, including wind speed, wind direction, temperature structure, and the presence or absence of effective acoustic ducts. At short source-receiver distances, bolide infrasound may be observed as direct arrivals, typically within $<$250 kilometers of the source. At larger distances, detection generally requires refraction through atmospheric waveguides, where wind and temperature structure return acoustic energy toward the surface. Thus, atmospheric propagation strongly influences whether a signal from a given event is observed at a particular station. Numerous studies have examined bolide-generated infrasound using seismic, acoustic, optical, or combined datasets, often focusing on individual events or limited samples selected for specific scientific or hazard-driven objectives \citep{Arrowsmith2008, Brown2011, McFadden2021, Wilson2025}. While these efforts established critical baselines, they have generally relied on small event samples, heterogeneous detection strategies, or regionally constrained observations, limiting population-level assessment of the physical factors governing long-range infrasound detectability.

Global observations of infrasound are enabled by the International Monitoring System (IMS), operated by the Preparatory Commission for the Comprehensive Nuclear-Test-Ban Treaty Organization \citep{Christie2010}. The IMS comprises a worldwide network of infrasound arrays designed primarily to detect nuclear explosions for treaty verification, but it has also proven highly effective for recording energetic natural phenomena, including bolides. Because these sensors operate continuously and independently of optical conditions, they provide a unique capability for monitoring atmospheric energy deposition on a global scale. Complementary information on energetic meteoroid entry events is provided by the fireball database maintained by the National Aeronautics and Space Administration (NASA) Jet Propulsion Laboratory (JPL) Center for Near-Earth Object Studies (CNEOS). This database is compiled from United States Government space-based sensor detections and includes associated metadata \citep{Tagliaferri1994}. Owing to the global coverage of these sensors and their relatively high energy-detection threshold (typically corresponding to events with impact energies on the order of $\sim$0.1 kt TNT equivalent or greater), the CNEOS database captures the most energetic bolide entries, those most likely to generate detectable infrasound, and has become a cornerstone dataset for studies of atmospheric entry physics and planetary defense \citep[e.g.,][]{Brown2013, Pilger2015, Pilger2020}.

Previous studies have analyzed numerous fireball events using seismic and acoustic datasets to assess ground impacts and implications for public safety \citep{Gi2018, Neidhart2021}. These efforts have generally focused on high-yield events, typically exceeding 1 kiloton (kt) of trinitrotoluene (TNT) equivalent (1 kt TNT = 1.485×10\textsuperscript{12} J), resulting in comparatively small samples of roughly 20 well-documented cases. Expanding this scope, a recent study conducted a statistical analysis of more than 100 events to distinguish large bolides from smaller meteors using flight parameters and total known mass as primary discriminators \citep{Betzler2025}. Similarly, another study compiled a database of 71 bolides to investigate the infrasound period–yield relationship, finding that global-scale detections were generally associated with bolide energies above 20 kt \citep{Ens2012}. Their work also demonstrated that, because infrasound propagation is strongly influenced by time-varying atmospheric conditions, range-dependent atmospheric models are essential for accurate source–receiver characterization. A follow-on study expanded the dataset, further refining the period–yield relationship \citep{Gi2017}. More recently, a statistical analysis of numerous global bolide events demonstrated that period–yield relationships are more nuanced than previously recognized and are strongly influenced by entry parameters and fragmentation style \citep{Silber2025c}. A complementary approach calibrated an ablation model using 59 Earth-impacting fireballs, incorporating orbital and physical strength estimates to improve source characterization \citep{Brown2016}.

In this study, we conduct a systematic global search for infrasound signals associated with all CNEOS-reported bolide events occurring between 22 September 2007 and 31 May 2025, leveraging the full IMS infrasound network and a uniform, automated detection and processing framework. By analyzing both bolides that generate detectable infrasound and those that do not, we examine how meteoroid entry parameters and atmospheric propagation conditions control infrasound detectability at regional to global scales. This approach yields the largest compiled dataset of bolide-generated infrasound detections to date and enables population-level analyses that were not previously possible.

By comparing the properties of infrasound-detected bolides with those of the broader CNEOS population, this work provides new constraints on the physical and environmental factors that govern atmospheric energy deposition and long-range acoustic propagation. These results advance the use of infrasound as a quantitative tool for planetary defense, atmospheric science, and global monitoring, and establish a foundation for future studies that integrate bolide entry physics with global acoustic observations.

This paper is organized as follows: Section 2 summarizes meteor-generated shock waves and the formation of infrasound, Section 3 describes the methods, Section 4 presents the results, and Sections 5 and 6 provide the discussion and conclusions, respectively.

\section{Background and Theory}\label{background}

To provide a physical framework for interpreting the statistical trends observed in the global dataset, we first review the fundamental mechanisms governing bolide shock generation and the subsequent coupling of this energy into long-range atmospheric waveguides. In particular, understanding the distinction between ballistic and fragmentation shock geometries, as well as the environmental controls on propagation, is essential for identifying the factors that govern global infrasound detectability.

\subsection{Shock Generation During Meteoroid Atmospheric Entry}\label{shock_generation}

When a meteoroid enters Earth's atmosphere at hypersonic velocities, it generates a strong shock wave \citep{ReVelle1976, Bronshten1983, Ceplecha1998, Silber2018}. During the hypersonic flight phase, while the meteoroid remains largely coherent but undergoes continuous ablation, energy is deposited quasi-continuously along the atmospheric path. Under these conditions, the resulting shock is commonly approximated as a cylindrical line source, with pressure disturbances radiating predominantly perpendicular to the direction of motion \citep{ReVelle1976, Silber2018}. The geometry of the meteoroid trajectory plays a key role in determining how and where this shock energy is deposited. Entry angle influences the rate at which the meteoroid encounters denser atmospheric layers and therefore controls the vertical extent of energy release along the path: steeper trajectories tend to concentrate energy deposition over shorter path lengths, whereas shallower entries distribute energy over longer distances \citep{Brown2013, Popova2013, Silber2025c}. These geometric differences directly affect the spatial distribution of shock generation and the effective source geometry perceived by distant observers \citep{Pilger2015}.

Fragmentation further modifies this picture by introducing localized regions of enhanced energy release along the trajectory, which can generate quasi-spherical shock components superimposed on the underlying cylindrical line-source shock. Although the detailed physics of fragmentation is complex and event-specific, its occurrence alters the spatial and temporal structure of shock generation by superimposing discrete or distributed source regions onto the underlying hypersonic flight path \citep{Bronshten1995, Borovicka1998, Ceplecha1998, Borovicka2020, Trigo2021}. As a result, bolide shock production cannot, in general, be described by a single idealized source, but rather by a combination of extended and localized contributions whose relative importance varies from event to event. As the shock propagates away from the source region, its initially nonlinear behavior weakens due to geometric spreading and dissipative processes \citep{Zeldovich2002}. At sufficiently large distances from the trajectory, the disturbance transitions into a linear or weakly nonlinear acoustic wave, allowing energy to radiate away from the source over large spatial scales \citep{ReVelle1976}.

\subsection{Transition to Infrasound and Long-Range Propagation}\label{infrasound}

The low-frequency component of the acoustic radiation produced during bolide entry is referred to as infrasound and occupies frequencies below 20 Hz \citep{Evans1972}. Infrasound experiences relatively minor atmospheric attenuation compared to higher-frequency acoustic waves, allowing it to persist over regional to global distances under favorable conditions \citep{ReVelle1976, Drob2003, Marcillo2013}. Consequently, infrasound provides a powerful means of observing atmospheric energy deposition by bolides far beyond the range of optical or regional sensors.

The detectability of bolide-generated infrasound depends on both the characteristics of the source and the atmospheric environment through which the acoustic energy propagates. Source geometry, governed by entry angle and the altitude range over which shock generation occurs, influences the spatial distribution and coherence of the acoustic radiation. Atmospheric structure then determines whether this radiation is refracted back toward the surface and transmitted over long distances \citep{Sutherland2004}. Long-range infrasound propagation is strongly controlled by vertical gradients in temperature and horizontal winds, which can form atmospheric waveguides, commonly referred to as ducts, in the troposphere, stratosphere, and thermosphere \citep{Waxler2015, Albert2023}. The presence and strength of these ducts can be assessed using the effective sound speed, defined as the sum of the adiabatic sound speed and the wind component along the propagation direction \citep{Godin2002}. When the effective sound speed aloft exceeds that near the surface, acoustic energy may be refracted back toward the ground, increasing the likelihood of detection at large source-receiver distances.

As a result, the observation of bolide-generated infrasound reflects an interplay between meteoroid entry geometry and atmospheric propagation conditions. Even highly energetic events may fail to produce detectable infrasound if propagation conditions are unfavorable, while events with more modest energies can be detected at regional to global distances when atmospheric ducting is strong \citep{Bowman2025}. To empirically constrain these dependencies, we now turn to a systematic global analysis combining satellite-derived bolide trajectories with long-term infrasound network observations.

\section{Methods}\label{methods}

\subsection{Bolide Dataset}\label{bolide_dataset}

This study is based on bolide events reported in the CNEOS fireball database, which provides globally documented detections of energetic meteoroid atmospheric entry events recorded by United States Government sensors. The catalog includes event timing (UTC), a single reported geographic location, peak brightness altitude, estimated impact energy, and, for many events, velocity vector components, with energy reported in kilotons of TNT equivalent \citep[e.g.,][]{Silber2025d}. The reported location corresponds to an event-level source point rather than a full time-resolved trajectory, which introduces uncertainty when associating distributed acoustic arrivals with different portions of the bolide path. These parameters provide a consistent global dataset suitable for systematic investigation of bolide infrasound detectability. Since this study is anchored to the CNEOS fireball database, the reported detection fraction applies only to CNEOS-reported events. We do not attempt to identify bolide-like infrasound events without corresponding public CNEOS entries, and therefore the results should not be interpreted as a completeness estimate for the full global bolide population or for all bolide-generated infrasound. We also used the reported velocity components to derive entry angles following \citet{Pena2022}.

An advantage of the CNEOS database is its relatively high lower threshold of detectable energy deposition, which naturally biases the catalog toward events capable of generating strong shock waves and measurable acoustic signals. However, several important limitations are also recognized. CNEOS-reported parameters generally lack formal uncertainty estimates, and multiple studies have noted potential biases or reduced precision in certain quantities, particularly velocity vectors and derived orbital parameters \citep{Devillepoix2019, Brown2023, Pena2025}. In the present study, these limitations do not materially affect our analysis, as we focus primarily on statistical detectability, relative comparisons between detected and non-detected populations, and aggregate trends rather than precise reconstruction of individual trajectories.

The CNEOS catalog has been widely used in the literature as a benchmark dataset for studying bolide energetics, atmospheric entry processes, and infrasound generation \citep{Ens2012, Gi2017, Silber2025c, Silber2025d}. Impact energies reported in the database span from sub-kiloton events to tens of kilotons of TNT equivalent, with rare extreme cases such as the 2013 Chelyabinsk superbolide reaching nearly 0.5 megatons \citep{Brown2013}. This energy range overlaps well with the regime in which bolides are capable of producing detectable long-range infrasound, making CNEOS particularly well suited for global infrasound-based analyses.

\subsection{Event Selection and Infrasound Signal Search}\label{event_selection}

To identify infrasound signals associated with CNEOS-reported bolide events, we conducted a systematic search across the global IMS infrasound network. Because consistent waveform availability is required for automated array processing, our analysis is restricted to events occurring between 22 September 2007 and 31 May 2025, corresponding to the period for which sufficiently complete IMS waveform data are available. Within this interval, the CNEOS database reports 623 bolide events, which form the baseline population examined in this study.

Predicted tropospheric and thermospheric apparent celerities, derived using the CNEOS-reported event locations, were used to define the initial arrival-time windows for waveform analysis \citep{Negraru2010, Silber2024}. However, the CNEOS ground truth does not necessarily correspond to the precise time or location of dominant infrasound generation along the bolide trajectory. To account for uncertainty in source timing, spatial extent, and propagation-path geometry, we therefore adopt an expanded search window spanning apparent celerities from 150 to 600 m/s, which is broader than the typical infrasonic celerity range. This ensures that early or delayed arrivals associated with distributed source regions or complex propagation paths are not excluded \textit{a priori}.

To further refine detections and reduce the number of low-probability event–station combinations, we applied a distance–yield gating scheme following established practice. IMS stations located more than 10,000 km from events with estimated impact energies below 1 kt were excluded from the search \citep{Ens2012, Silber2025c}. For events with energies between 1-2 kt, a maximum source-receiver distance of 15,000 km was imposed, whereas all available IMS stations were considered for events exceeding 2 kt. This strategy prioritizes physically plausible detection scenarios while substantially reducing computational overhead associated with extremely low-likelihood event-station pairings.

\subsection{Signal Analysis}\label{signal_analysis}

Waveform data extracted within the predefined arrival-time windows were analyzed using beamforming, a well-established array processing technique that exploits the relative arrival-time delays of an incident planar wavefront across multiple sensors. By coherently shifting and summing the recordings, beamforming amplifies coherent signal energy while suppressing incoherent noise \citep{Rost2002, Rost2009}. In this study, beamforming was implemented in the frequency domain to estimate back azimuth (direction of arrival) and trace velocity via grid search. This frequency-domain implementation, commonly referred to as frequency-wavenumber (\textit{fk}) analysis, may be applied using either a single frequency band or multiple bands \citep{Capon1969}.

Single-band processing assumes that the signal of interest is fully contained within the selected frequency range, an assumption that may not hold for bolide-generated infrasound due to spectral variability. In addition, single-band approaches do not account for temporal variations in noise spectral content, which may vary across frequencies during the analysis window and further obscure the target signal. For these reasons, multi-band \textit{fk} analysis is often advantageous in complex noise environments.

Processing was performed using Cardinal, an open-source, multi-frequency array processing framework designed for systematic detection and characterization of coherent acoustic arrivals across infrasound arrays of various apertures \citep{Ronac2025}. Cardinal implements \textit{fk} processing across each pixel in a segmented time-frequency space, enhancing estimation of signal directionality, trace velocity, and waveform coherence. While the software supports multiple band-spacing configurations, we adopt a one-third-octave spacing to enable detailed characterization of bolide-generated infrasound \citep{Iezzi2022}.

Cardinal comprises four primary components: (1) the Segmentor, which discretizes time-frequency space; (2) the Adaptive Array, which determines optimal subarray geometry for multi-scale arrays across multiple frequency bands to maximize coherence; (3) the Array Processor, which applies \textit{fk} processing to each time-frequency pixel; and (4) the Aggregator, which groups coherent detections into families based on similarity in arrival parameters \citep{Ronac2025}. The framework also includes a preprocessing module that evaluates station state-of-health, identifies polarity inversions, and quantifies data gaps. In this study, waveform data were automatically processed using the first three components, while the preprocessing module and Aggregator were applied selectively to further enhance weak detections. The discretization scheme used throughout this study is summarized in Table A1. Additionally, the Cardinal preprocessing workflow provided quality-control information on station state-of-health, polarity inversions, and data gaps, which was used to identify problematic waveform segments during the detection review. However, we did not apply a formal station-by-station observing-efficiency correction or time-dependent noise completeness model; therefore, some non-detections may reflect unfavorable local noise conditions or incomplete usable waveform availability rather than an absence of acoustic energy from the source.

A valid detection is defined as a signal whose back azimuth is consistent with the CNEOS-reported event location, whose celerity lies within the predicted infrasound range of 180-330 m/s \citep{Negraru2010}, and whose trace velocity falls within the expected infrasonic range of 250-500 m/s \citep{Lonzaga2015}. These criteria follow established practice for associating infrasound detections with known sources \citep{Arrowsmith2008} and allow reliable discrimination between bolide-generated signals and unrelated atmospheric or anthropogenic noise, while retaining sensitivity to broadband, temporally extended, or multi-phase arrivals characteristic of bolide infrasound.

\subsection{Atmospheric Conditions and Propagation Environment}\label{atmospheric_conditions}

Because atmospheric structure exerts a first-order control on long-range infrasound propagation, meteorological conditions along each bolide-station path were characterized using the Ground-to-Space (G2S) atmospheric model \citep{Drob2003}. G2S is a semi-empirical, physics-based atmospheric specification that combines data-assimilative lower-atmosphere models with empirical representations of the middle and upper atmosphere to produce vertically continuous profiles of temperature and horizontal winds from the surface to thermospheric altitudes. The model is widely used in infrasound and seismoacoustic studies to characterize large-scale propagation environments under realistic atmospheric conditions. Given the large number of source-receiver paths evaluated in this study, we used the G2S command-line client to automate retrieval and storage of all required parameters \citep{Hetzer2024}.

For each event-station pair, vertical profiles of temperature and horizontal winds were extracted from G2S at the midpoint of the great-circle path between the CNEOS-reported bolide location and the receiving IMS station. This midpoint approximation provides a consistent means of sampling the propagation environment across the large number of source-receiver paths considered in this study. The spatial relationship between bolide locations and IMS stations, including representative path geometries, is illustrated in Figure A1.

From the G2S profiles, the effective sound speed as a function of altitude was computed by combining the adiabatic sound speed with the horizontal wind component projected along the propagation direction \citep{Godin2002}. These profiles were then used to assess the presence of atmospheric ducting conditions capable of supporting long-range infrasound propagation, including tropospheric, stratospheric, and thermospheric waveguides. Atmospheric waveguides form when vertical gradients in temperature and horizontal winds refract acoustic energy back toward the surface, enabling long-range transmission. In this study, we identify favorable ducting conditions using positive vertical gradients in effective sound speed, which indicate refractive conditions favorable for returning acoustic energy toward the surface.

Midpoint atmospheric profiles provide a practical first-order representation of the propagation environment across many source-receiver paths and enable consistent comparison of atmospheric conditions among events. While this approach does not capture full range-dependent variability, such as horizontal gradients or temporal evolution along the propagation path, it is sufficient for comparative analysis of detectability across the global dataset considered here. Accordingly, atmospheric profiles are used to contextualize observed detections rather than to perform detailed ray-tracing or full-wave propagation modeling.

\subsection{Population-Specific Analysis}\label{population_specific_analysis}

To compare the distributions of bolide flight parameters between the full CNEOS catalog and the subset of events that produced infrasound detections, we computed kernel density estimates (KDEs) for entry angle, entry velocity, and altitude. KDEs provide a nonparametric estimate of the underlying probability density function by placing a Gaussian kernel at each data point and summing the contributions to produce a smooth continuous distribution, without assuming any specific analytic form. The bandwidth (smoothing parameter) was selected automatically using a variant of Scott's Rule, ensuring that the KDE adapts to the spread and variance of the input data \citep{Scott1992, Virtanen2020}. By evaluating the KDE on a dense grid spanning the range of observed values, we obtain continuous probability density functions suitable for comparing the statistical structure of the full CNEOS population with that of the infrasound-detected subset.

To quantify how the flight characteristics of bolides associated with infrasound detections differ from the broader CNEOS population, we estimate the underlying probability density functions for entry angle, peak brightness altitude, and entry velocity using the KDE approach described above. While KDEs provide a smooth, intuitive view of how bolide flight parameters are distributed, they can sometimes obscure differences between populations, particularly when those differences occur in the distribution tails or appear as systematic shifts rather than local density peaks. To more clearly assess how the detected and non-detected populations diverge across the full parameter range, we also compute empirical cumulative distribution functions (CDFs). Unlike KDEs, CDFs quantify the cumulative probability that a parameter is less than or equal to a given value, making them particularly sensitive to systematic shifts between distributions.

To quantify relationships among bolide entry parameters, we compute Pearson correlation coefficients as the primary measure of association presented in the main text. Pearson correlations evaluate linear dependence between variables and therefore emphasize systematic directional trends in parameter space. In our analysis, this metric is particularly useful for highlighting the inverse relationship observed between entry angle and entry velocity when comparing the subset of bolides that produced detectable infrasound with the broader CNEOS population. However, Pearson correlations can be influenced by outliers and may underrepresent nonlinear but monotonic relationships. To assess the robustness of the observed trends, we therefore also compute Spearman rank-order correlation coefficients, which measure monotonic association without assuming linearity and are less sensitive to non-Gaussian structure. These results are provided to complement the Pearson analysis and to ensure that our interpretations do not depend on the choice of correlation metric. Lastly, since entry angle is derived from velocity components, and velocity information is not uniformly available across the CNEOS catalog, we also assess whether the subset of events with reported velocity vectors is representative of the broader population. This allows us to evaluate whether observed trends in entry angle could arise from selection biases within the velocity-resolved subset.

\section{Results}\label{results}

\subsection{Infrasound Detection Statistics}\label{infrasound_detection_statistics}

Our systematic search includes 623 CNEOS bolide events occurring between 22 September 2007 and 31 May 2025, spanning a wide range of entry velocities, trajectory geometries, and estimated impact energies. Within this population, we identified 311 distinct bolide events that produced verifiable infrasound signals. This corresponds to an overall infrasound detection rate of roughly 50\% relative to the full CNEOS catalog. These confirmed events yielded a total of 573 individual infrasound detections distributed across the global IMS network, reflecting the capability of the network to record single events at multiple stations. Of the 623 bolides analyzed, 333 events include velocity vector information to compute entry angles. Accounting for multiple array detections per event, this yields 407 detections with associated entry angle estimates.

\subsection{Example Detection and Signal Characterization}\label{example_detection}

Figure 1 illustrates representative Cardinal array processing results for a bolide event that occurred on 26 June 2022 (Figure A1). The processed waveforms exhibit a high signal-to-noise-ratio (SNR) arrival with a stable back azimuth that is consistent with the CNEOS-reported source location. The observed trace velocities fall within the expected range for atmospheric infrasound, and the semblance panel demonstrates strong signal coherence across the array. Back azimuth consistency was used as one of the criteria for identifying valid bolide infrasound detections. Candidate arrivals were retained when the observed back azimuth was within $\pm$15$^\circ$ of the predicted value from the CNEOS-reported source location to the station. For very short source-receiver distances ($\sim$100 km), a wider tolerance of $\pm$45$^\circ$ was applied to account for increased sensitivity of azimuth to small spatial offsets. The retained detections therefore show overall azimuthal consistency with the reported CNEOS event locations, although some scatter is expected because bolide acoustic energy may be generated along an extended trajectory rather than at a single point. These characteristics confirm a coherent infrasound arrival associated with the 26 June 2022 bolide event. The optimal subarray geometry selected for each processing frequency band is shown in Figure A2.

\begin{figure}[htpb]
\centering
\includegraphics[width=\textwidth]{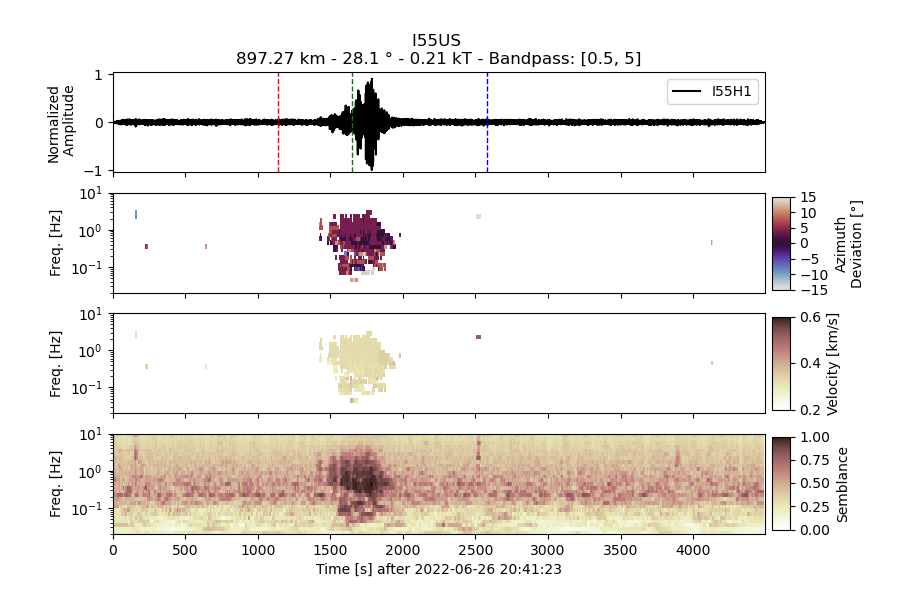}
\caption{Infrasound array processing results for a bolide that occurred on 26 June 2022. The top panel shows the bandpass-filtered (0.5-5 Hz) waveform from a single array element, with dashed vertical lines marking the predicted arrival times for tropospheric (red), stratospheric (green), and thermospheric (blue) phases. The second panel displays back azimuth estimates expressed as deviations from the CNEOS-reported source direction. The third panel shows the corresponding trace velocity estimates, and the bottom panel presents the semblance values, where a threshold of 0.5 was applied to visualize coherent arrivals. Optimal subarray geometry for each frequency band used for processing is shown in Figure A2}
\label{fig:1}
\end{figure}

Atmospheric conditions associated with this detection were characterized using effective-sound-speed profiles derived from G2S midpoint atmospheric specifications, as described in Section 3.4. These profiles are used to contextualize detections rather than to predict detailed propagation paths. Figure A1 illustrates the spatial relationship between the reported bolide location and the global IMS network, while Figure A3 presents representative temperature, wind, and effective-sound-speed profiles for the 26 June 2022 event shown in Figure 1. For this case, the effective-sound-speed profile exhibits a positive vertical gradient near 40 km altitude, indicative of a stratospheric duct along the propagation path between source and receiver (I55US). Such a gradient enables upward-propagating acoustic energy to refract back toward the surface rather than leaking into the upper atmosphere. This ducting regime is consistent with the stratospheric arrival identified in the array processing results and provides a physically plausible explanation for why the event was detectable at I55US.

This example demonstrates the effectiveness of the multi-frequency array processing approach in isolating coherent infrasound arrivals from background noise and illustrates the types of signals included in the broader statistical analysis.

\subsection{Global Distribution}\label{global_distribution}

Figure 2 shows the global distribution of all CNEOS-reported bolides together with the subset of events that produced detectable infrasound. The full catalog exhibits broad geographic coverage, with events occurring across nearly all latitudes and longitudes, reflecting the global nature of near-Earth object entry. The subset of bolides with infrasound detections is similarly widespread, indicating that detectable events are not confined to any particular geographic region. Instead, the spatial pattern closely mirrors the underlying bolide distribution, suggesting that infrasound detectability is governed primarily by atmospheric propagation conditions and station geometry rather than by regional observational bias. The presence of detections over both oceanic and continental regions, despite the uneven global distribution of infrasound arrays, further highlights the efficiency of long-range acoustic propagation in capturing energetic bolide events worldwide.

\begin{figure}[htpb]
\centering
\includegraphics[width=\textwidth]{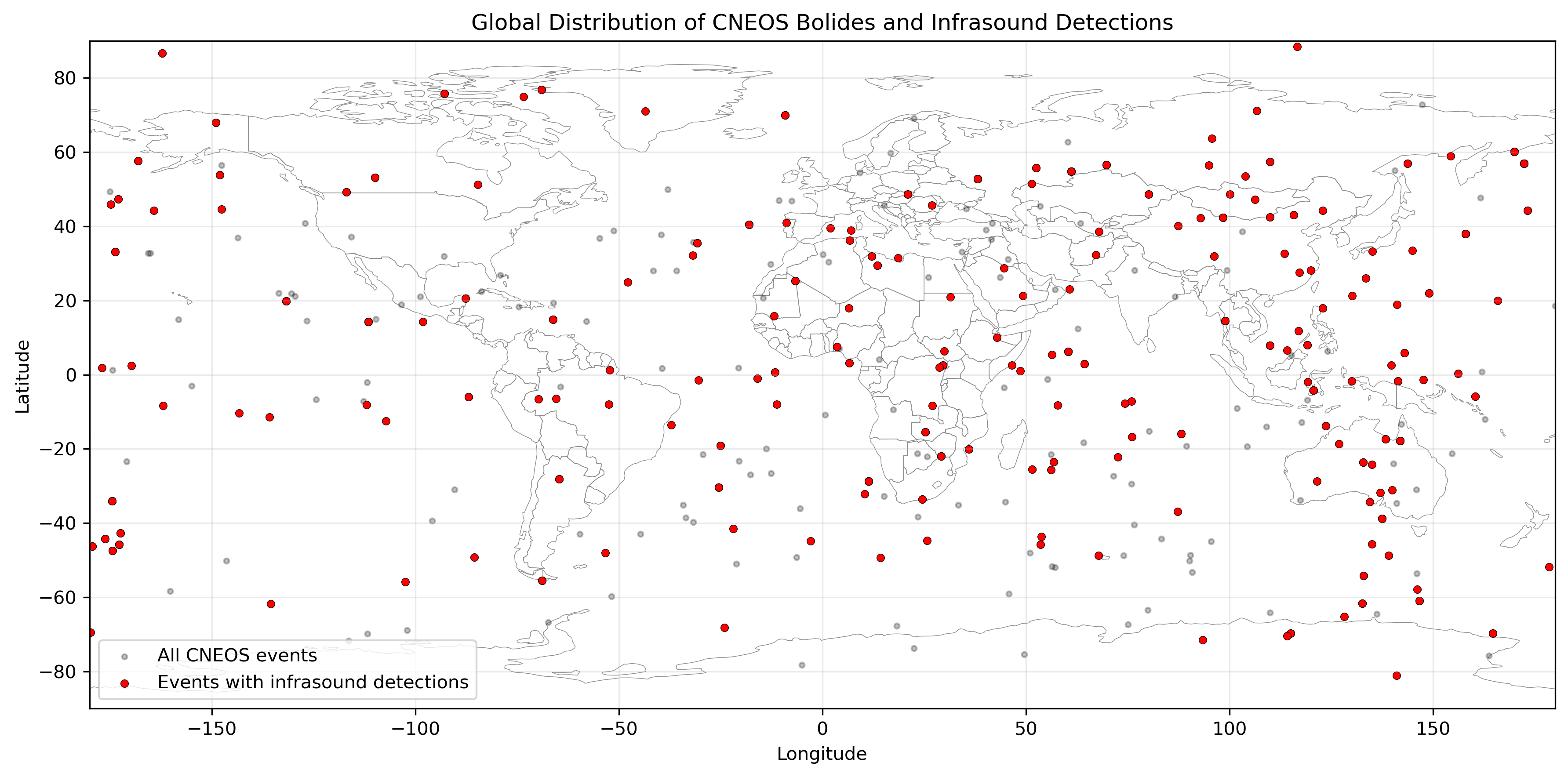}
\caption{Global distribution of CNEOS-reported bolides (gray) and the subset that produced detectable infrasound (red). Bolides occur worldwide, and the spatial distribution of detected events closely mirrors the full catalog, reflecting the global reach of the infrasound network and the efficiency of long-range acoustic propagation. Global distribution of IMS stations can be seen in Figure A1.}
\label{fig:2}
\end{figure}

\subsection{Distributions of Bolide Entry Parameters}\label{entry_parameters}

To examine how bolide entry characteristics differ between the full CNEOS population and the subset of events that produced detectable infrasound, we compare the distributions of entry angle, entry velocity, and peak brightness altitude. Distributional comparisons are performed using KDEs and empirical CDFs, as described in Section 3.5. Figure 3 shows KDEs for the three entry parameters. Among these, entry angle exhibits the most pronounced difference between detected and non-detected events: bolides associated with infrasound detections occur more frequently at steeper entry angles than those in the full CNEOS catalog. Peak brightness altitude also displays a systematic shift, with detected events tending to reach peak brightness at lower altitudes relative to the broader population. In contrast, the entry-velocity distributions for detected and non-detected bolides are broadly similar, although the detected subset shows a modest shift toward lower velocities.

\begin{figure}[htpb]
\centering
\includegraphics[width=0.4\textwidth]{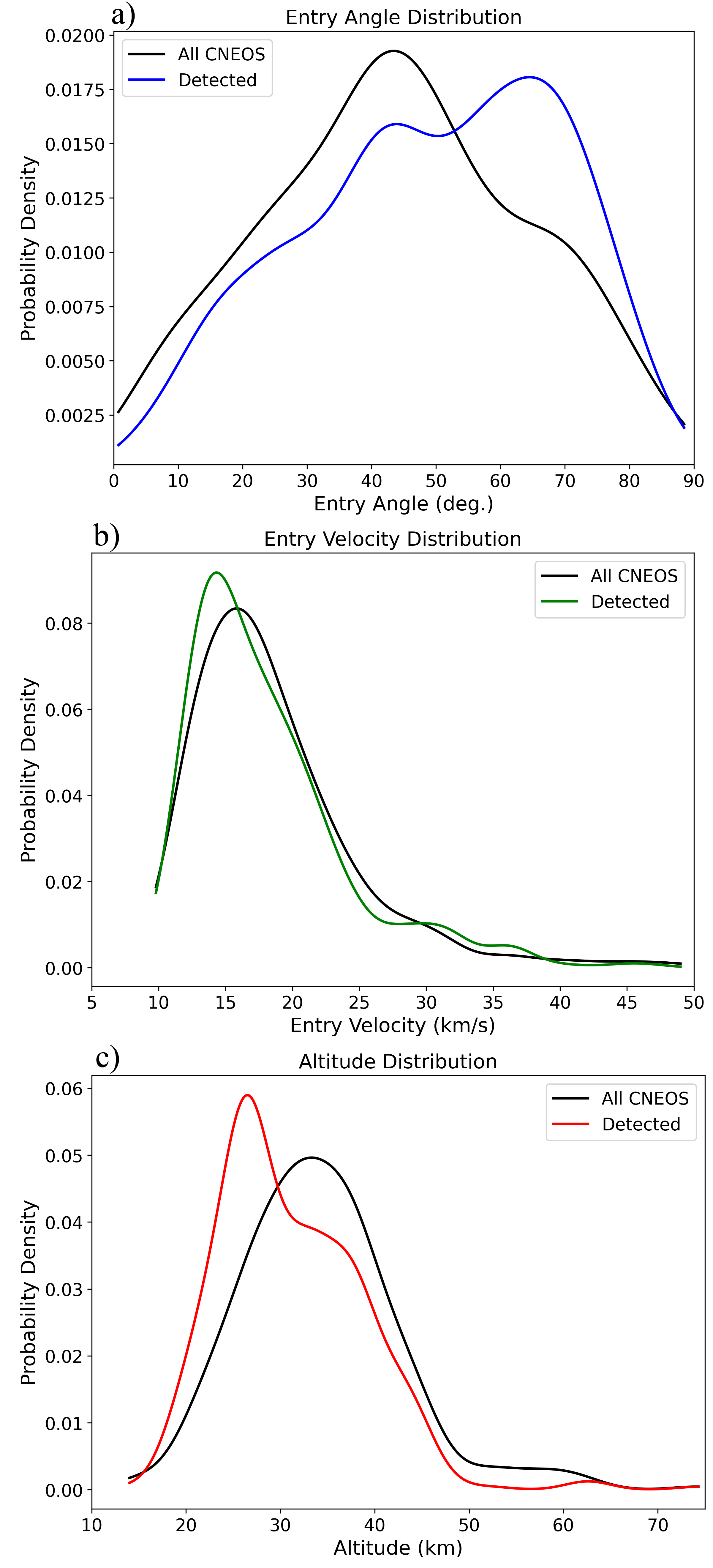}
\caption{Kernel density estimates (KDEs) comparing the distributions of flight parameters for all CNEOS-reported bolides (black curves) and those that produced infrasound detections (colored curves). Panel (a) shows entry angle distributions and an enhanced density at steeper entry angles among detected events. Panel (b) displays the entry velocity distributions and how they are broadly similar between the two populations, with a minor shift toward slower velocities for detected bolides. Panel (c) plots the peak brightness altitude distributions, with detected events favoring slightly lower altitudes relative to the full catalog. }
\label{fig:3}
\end{figure}

While KDEs provide a smooth visualization of distributional structure, systematic differences between populations can be more clearly assessed using CDFs. Because the strongest differences appear in entry angle and altitude, we restrict our CDF analysis to these two parameters. The resulting empirical cumulative distribution functions are shown in Figure 4.

\begin{figure}[htpb]
\centering
\includegraphics[width=0.5\textwidth]{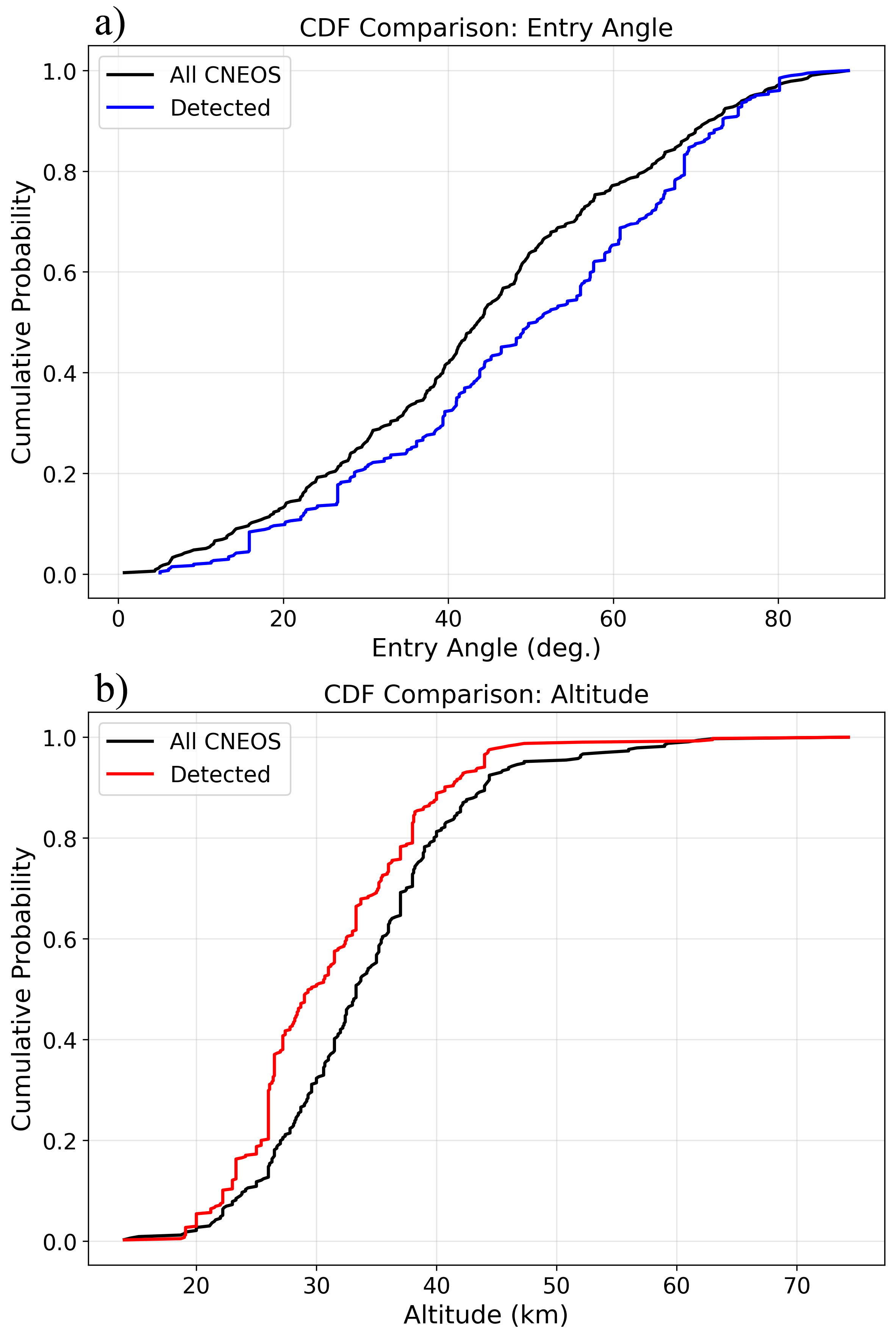}
\caption{Cumulative distribution functions (CDFs) comparing the full CNEOS bolide catalog (black lines) with the subset of events that produced infrasound detections (colored lines). Panel (a) plots the entry angle CDFs while panel (b) shows the altitude CDFs.}
\label{fig:4}
\end{figure}

For entry angle, the CDF corresponding to infrasound-detected events lies consistently below that of the full CNEOS population, indicating that detected bolides reach a given cumulative probability at steeper angles. For peak brightness altitude, the detected-event CDF lies above the full-population curve, demonstrating that detected bolides tend to reach peak brightness at lower altitudes. These shifts persist across the full parameter range and are not confined to the central portions of the distributions.

Taken together, the KDE and CDF results show that bolides producing detectable infrasound differ systematically from the broader CNEOS population in terms of entry geometry and altitude of peak brightness, whereas entry velocity alone does not strongly distinguish the two populations.

\subsection{Correlations Among Entry Parameters}\label{correlations}

To examine how bolide entry parameters co-vary within the analyzed populations, we compute Pearson correlation coefficients for entry angle, entry velocity, and peak brightness altitude. The resulting correlation matrices for the full CNEOS population and for the subset of events that produced detectable infrasound are shown in Figure 5. For the full CNEOS catalog, correlations among the three parameters are weak, indicating that entry angle, velocity, and altitude are only loosely coupled across the global bolide population.

\begin{figure}[htpb]
\centering
\includegraphics[width=0.5\textwidth]{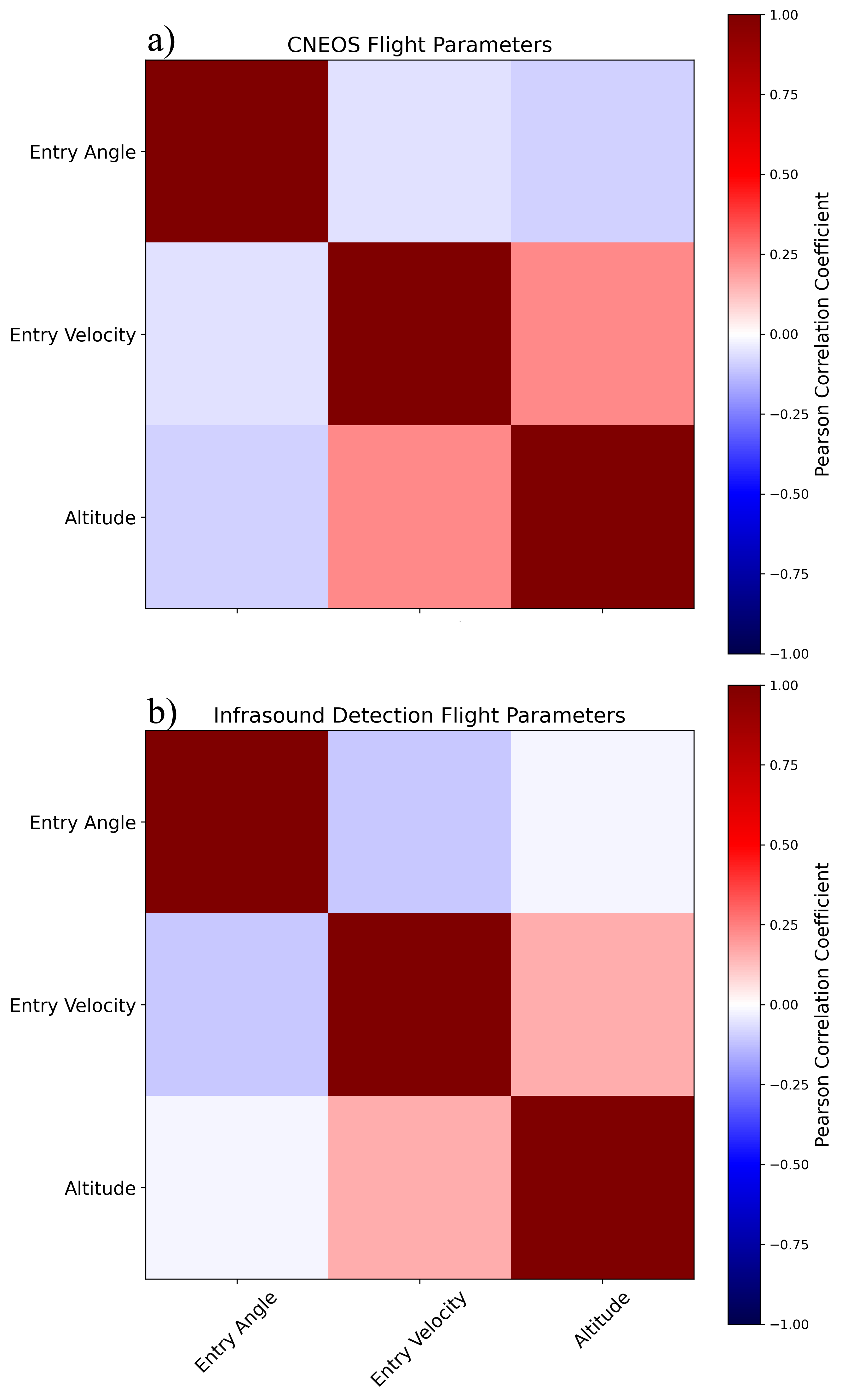}
\caption{Pearson correlation matrices for bolide entry parameters. Panel (a) shows correlations among entry angle, entry velocity, and peak brightness altitude for all CNEOS-reported bolides. Panel (b) plots the correlations for the subset of events that produced infrasound detections. The color scale indicates Pearson correlation coefficients from -1 to 1.}
\label{fig:5}
\end{figure}

In contrast, the subset of bolides that produced detectable infrasound exhibits a more pronounced correlation structure. Entry angle and entry velocity show a moderate inverse relationship, such that detected events arriving at steeper angles tend to be associated with lower velocities. Peak brightness altitude is also positively correlated with entry velocity within the detected subset. While weak correlations are present in the full CNEOS population, they are notably stronger in the detected subset, indicating that the covariance among entry parameters becomes more structured within the subset of events that produce detectable infrasound.

Since entry angle is computed only for events with available velocity vectors, we evaluate whether this subset is representative of the broader CNEOS database. Statistical comparisons indicate that the velocity-resolved subset is modestly biased toward higher-energy and slightly lower-altitude events (Figure A4). To further assess whether these biases influence detectability trends, we compare the distributions of flight parameters within the velocity-resolved subset between detected and non-detected events (Figure A5). Within this subset, detected bolides remain systematically shifted toward steeper entry angles relative to non-detected events, while entry velocity distributions are nearly indistinguishable between the two populations, indicating that entry velocity is not a primary controlling factor for infrasound detectability within this population. In contrast, peak brightness altitude distributions show substantial overlap between detected and non-detected events, suggesting that altitude plays a secondary role in detectability and is more strongly influenced by underlying population biases than entry angle. These results indicate that the observed dependence of infrasound detectability on entry angle is robust to selection effects associated with velocity-vector availability. Entry angle is therefore the dominant discriminator of detectability, while altitude plays a secondary role and appears more sensitive to population-level sampling biases than entry angle.

Due to the fact that Pearson correlations emphasize linear dependence and can be influenced by distributional structure, we also compute Spearman rank-order correlations as a complementary, nonparametric measure of monotonic association (Figure A6). While the Spearman coefficients show the same directional relationships, the contrasts between the detected and full populations are weaker than in the Pearson analysis. This indicates that the observed covariance reflects a combination of linear structure and substantial scatter, rather than a strongly ordered monotonic relationship across the full parameter range.

\subsection{Influence of Bolide Energy on Detectability}

To evaluate whether the observed dependence of infrasound detectability on entry geometry could be driven by event energetics, we examine the distribution of bolide energies for detected and non-detected populations. Since higher-energy events are expected to produce stronger acoustic signals, they may be detectable regardless of entry geometry and could therefore introduce apparent geometric biases. Comparisons of energy distributions show that detected bolides are, on average, associated with higher reported energies, but that there is substantial overlap between detected and non-detected populations across the full energy range (Figure A7). This overlap demonstrates that energy alone does not uniquely determine detectability, particularly within the moderate-energy regime that comprises the majority of events. The detected population therefore represents a biased, but not exclusive, subset of the full CNEOS energy distribution.

To further assess whether the entry-angle dependence is an artifact of energy, we examine entry-angle distributions within comparable energy ranges. Within these subsets, detected events remain systematically shifted toward steeper entry angles across all energy bins, demonstrating that the relationship between entry angle and detectability persists independently of energy. Although the separation becomes less pronounced at the highest energies, these results confirm that while energy increases the overall likelihood of detection, it does not fully explain the observed geometric dependence. Source geometry therefore exerts a primary and independent control on detectability across the broader bolide population, with energy acting as a secondary modulating factor. This demonstrates that the observed dependence of infrasound detectability on entry angle is not an artifact of energy selection effects.

\subsection{Influence of Atmospheric Propagation on Detectability}

Finally, to assess whether atmospheric propagation effects could account for the observed detectability trends, we examine the relationship between entry geometry and acoustic waveguide strength. We also examined the relationship between entry angle, source altitude, and the strength of acoustic waveguides along source-receiver paths. Waveguide favorability was assessed using the vertical structure of effective sound speed, with positive vertical gradients indicating refractive conditions favorable for returning acoustic energy toward the surface. The results show substantial variability in ducting conditions across events and propagation paths, with no systematic relationship between entry angle and waveguide strength. This indicates that favorable propagation conditions are not preferentially associated with specific entry geometries. Instead, waveguide strength depends primarily on source altitude, path geometry, and the atmospheric temperature and wind structure sampled along each propagation path.

\section{Discussion}\label{discussion}

\subsection{Infrasound Detectability as a Selective Physical Process}\label{infrasound_detectability}

Early global studies of bolide-generated infrasound \citep{Ens2012, Gi2017}, implied a global detection fraction of <20\% when evaluated against the full CNEOS catalog through February 2016, reflecting both limited event samples and the evolving state of the IMS network. Conversely, our systematic analysis identifies infrasound detections for approximately 50\% of CNEOS bolides between 2007 and 2025, representing nearly a three-fold increase in observed detection rate relative to prior studies and reflecting expanded IMS coverage, improved waveform availability, and advances in detection methodology rather than a change in underlying bolide-atmosphere physics.

Since the earlier studies, the IMS infrasound network has reached a more mature global configuration with increased station density \citep{Hupe2022}, and part of the increased detection rate reflects methodological advances in signal identification \citep{Ronac2025}. In this work, detections were identified using Cardinal, a recently developed multi-frequency array processing framework that differs from earlier single-band or time-domain approaches utilized in earlier bolide studies. By evaluating coherence across multiple frequency bands and employing adaptive, frequency-dependent subarray selection, Cardinal improves sensitivity to broadband, temporally extended, and low-SNR bolide infrasound signals that may not be detectable using fixed-geometry processing alone.

These advances demonstrate that infrasound is a more effective global observable for energetic atmospheric entry events than previously recognized when modern processing is applied. At the same time, the absence of detectable infrasound for nearly 50\% of CNEOS bolides indicates that detectability is not guaranteed for all events. While local station noise levels inevitably mask some weak arrivals, the strong systematic correlations with entry parameters indicate that detectability is principally modulated by source geometry and atmospheric propagation, with the atmosphere acting as a selective physical filter that favors those acoustic signals that couple efficiently into persistent waveguides. The non-detected population may also include false negatives, including events that were energetically or geometrically favorable but lacked confirmed infrasound detections. Such cases may reflect unfavorable source-receiver geometry, weak or absent ducting along available paths, elevated station noise, data gaps, temporary array performance issues, or offsets between the dominant acoustic source region and the CNEOS-reported peak-brightness location.

Accordingly, non-detection should not be interpreted solely as evidence that a bolide failed to generate infrasound. Because this study is intended as a broad population-level assessment, we do not examine matched detected and non-detected event pairs in detail. However, the similarity between the full and detected-event distributions indicates that detectability is not controlled by any single reported parameter; events with similar reported energies and broad entry geometries may still differ because of source-receiver distance, atmospheric propagation, station noise, and the detailed altitude-dependent structure of energy deposition \citep{Silber2025c}.

Source-receiver distance also affects detectability through geometric spreading, atmospheric attenuation, and path-dependent propagation. In this study, distance was incorporated through the distance-yield gating scheme described in Section 3.2, which limited the station search range for lower-energy events while retaining longer-range station pairs for higher-energy events. Although detectability generally decreases with distance, this dependence is not strictly monotonic because favorable atmospheric ducting can support efficient long-range propagation.

Due to the fact that the present analysis does not include a full station-level sensitivity or observing-efficiency model, some non-detections may instead reflect elevated local noise, temporary station performance issues, data gaps, or unfavorable source-receiver geometry rather than an absence of acoustic energy from the source. Quantifying these effects would require a formal observing-efficiency framework that combines station availability, array response, time-dependent noise, and propagation conditions, and is therefore left for future work. Overall, this population-level perspective provides important context for interpreting both detected and non-detected events and motivates the detailed examination of source geometry and atmospheric propagation that follows.

\subsection{Source Geometry and Entry Velocity}\label{source_geometry}

The results demonstrate that source geometry, rather than bolide energetics alone, exerts a primary control on infrasound detectability. In particular, bolides that enter at steep angles and deposit energy at lower altitudes are disproportionately over-represented among detected events. This behavior is consistent with the physics of meteor-generated shock waves, which during sustained hypersonic flight are well approximated as a cylindrical line source that continuously injects energy into the surrounding atmosphere along the trajectory \citep{ReVelle1976, Silber2018}. This interpretation is supported by the energy-controlled analysis presented in Section 4.6, which shows that the dependence of detectability on entry angle persists within comparable energy ranges, demonstrating that the observed geometric bias is not an artifact of energy selection effects but reflects a fundamentally geometry-driven detection process.

For steep entry trajectories, the hypersonic shock is generated over a shorter horizontal path length and a more vertically compact column \citep{Silber2024, Silber2025b}. In this configuration, acoustic energy is injected into the atmosphere at angles and altitudes that favor direct coupling into atmospheric waveguides, particularly the stratospheric duct \citep{Nippress2017, Albert2023, Bowman2025}. Since the ballistic shock front expands roughly perpendicular to the trajectory, steep-entry geometry promotes efficient redirection of acoustic energy into horizontally propagating modes that can be refracted back toward the surface and transmitted over long distances \citep{Albert2023}. Conversely, shallow-entry bolides distribute shock generation over much longer atmospheric paths. Although such events may deposit comparable or greater total energy \citep{Brown2013, Popova2013}, the resulting acoustic radiation is spatially distributed and injected at less favorable angles for waveguide capture \citep{Silber2025b, Silber2025c}. Different portions of the extended trail can contribute incoherently to the received signal, reducing effective coupling into a single dominant duct and diminishing long-range detectability. This geometric effect is consistent with previous analyses showing increased variability in back azimuths and apparent source locations for shallow-entry bolides, even under otherwise favorable propagation conditions \citep{Silber2025b}.

A moderate but systematic inverse relationship between entry angle and entry velocity emerges within the infrasound-detected population, indicating that detectability is modulated by a dual-constraint selection process rather than by intrinsic entry dynamics alone. In this interpretation, bolides are subject to an altitude constraint, whereby higher-velocity entries tend to deposit a larger fraction of their energy at higher altitudes due to rapid ablation and deceleration \citep{Popova1998, Popova2005}, and a source-waveguide coupling constraint, whereby shallow trajectories act as spatially extended sources that couple acoustic energy less coherently into atmospheric waveguides. Events that satisfy both constraints (low velocity and high entry angle) therefore tend to populate a narrow region of parameter space associated with improved global infrasound detectability, even though the relationship remains probabilistic rather than deterministic.

The observed positive correlation between peak brightness altitude and entry velocity among detected events further supports this interpretation. Faster bolides tend to deposit energy higher in the atmosphere \citep{Popova1998, Popova2005}, whereas slower and steeper entries more frequently concentrate energy deposition at altitudes favorable for sustained acoustic propagation. Importantly, the altitude of peak optical brightness does not necessarily correspond to the location of maximum acoustic energy generation. Optical luminosity reflects radiative processes, while infrasound generation is governed by the spatial and temporal distribution of shock energy coupled into the surrounding atmosphere \citep{ReVelle1976, Ceplecha1998}. For extended hypersonic trajectories well described by a cylindrical line-source approximation, acoustic radiation can originate from multiple regions along the path, with detectability dominated by those segments that are geometrically and atmospherically favorable for waveguide coupling \citep{Pilger2015, Silber2025b}. Peak brightness altitude therefore serves as a useful but imperfect proxy for acoustic source altitude, reinforcing that infrasound detectability depends on specific combinations of entry parameters, rather than on any single variable acting in isolation. Consistent with the statistical analysis, entry velocity alone does not act as a primary control on detectability, as velocity distributions are nearly indistinguishable between detected and non-detected events when controlling for velocity-vector availability, indicating that velocity does not independently govern infrasound observability.

Another likely contributor to detectability is the shape of the energy-deposition profile. Localized, burst-like energy release may couple more efficiently into long-range infrasound than flatter or more spatially extended deposition along the trajectory. Although light-curve morphology is not incorporated into the present population-level analysis, recent work has shown that CNEOS light curves can be systematically classified to characterize fragmentation and energy-deposition modes \citep{Silber2025d}, and that these classifications influence bolide infrasound energy relations \citep{Silber2025c}.

\subsection{Role of Atmospheric Propagation}\label{role_of_atmospheric_propagation}

While source geometry governs how and where acoustic energy is introduced into the atmosphere, infrasound detectability is further modulated by the propagation environment through which that energy travels \citep{Drob2003}. Infrasound propagation depends on vertical gradients in temperature and horizontal winds, which can form atmospheric waveguides that refract acoustic energy back toward the surface and enable long-range transmission. To evaluate whether the detection biases identified in this study are driven by systematic variations in atmospheric structure, we examine the relationship between entry geometry and the strength of the acoustic waveguide.

Specifically, we compare entry angle with band-averaged effective sound speed differences derived from G2S midpoint atmospheric specifications for stratospheric and thermospheric propagation regimes (Figure A8). This analysis reveals no significant correlation between entry angle and midpoint ducting strength across these altitude bands. The absence of a relationship indicates that the preferential detection of steep-entry bolides is unlikely to result from favorable atmospheric conditions coinciding with specific trajectories. Instead, entry geometry and atmospheric ducting strength appear to vary largely independently.

This finding supports the interpretation developed in Section 5.2 that improved detectability of steep-entry bolides arises primarily from source-waveguide coupling, with entry angle acting as the dominant geometric control on detectability. When ducting conditions are present, even if modest, steep-entry geometry increases the likelihood that acoustic energy is introduced into atmospheric waveguides in a manner conducive to long-range transmission. Shallow entries, by contrast, may fail to couple energy coherently into waveguides even under broadly similar atmospheric conditions.

These results should be interpreted with the limitation that back azimuth consistency was assessed relative to the CNEOS-reported source location. CNEOS provides a single event location and does not report formal location uncertainties, while bolide infrasound may originate from different portions of the trajectory. This is particularly relevant for shallow-entry events, for which geometric effects alone can produce measurable back azimuth deviations from a point-source prediction, even in the absence of atmospheric or measurement uncertainties.

The substantial scatter evident in Figure A8 further illustrates the limitations of range-independent (1D) atmospheric representations for global infrasound studies. Many bolide signals propagate over distances of several thousand kilometers, traversing regions with significant horizontal variability in wind and temperature. A single vertical profile extracted at the path midpoint cannot capture range-dependent effects associated with stratospheric warming events, tidal modes, gravity waves, or other mesoscale and synoptic-scale perturbations that influence acoustic propagation.

Consistent with this limitation, we observe clear detections even when midpoint profiles suggest weak or absent ducting, indicating that detectability is governed by path-integrated propagation conditions, rather than by a local atmospheric snapshot. The atmospheric analysis presented here should therefore be interpreted as a first-order characterization of propagation conditions rather than a full local-weather or range-dependent wind-field sensitivity study. Local and path-dependent variations in winds, temperature, turbulence, and station noise may further influence whether a given bolide signal is observed. While simple geometric acoustic metrics provide a useful first-order context for interpreting detectability, accurate prediction of global bolide infrasound observability is inherently probabilistic and would require full range-dependent modeling that accounts for the evolving four-dimensional structure of the atmosphere.

Across the dataset, we also identify a subset of infrasonic arrivals with apparent celerities faster than typically expected for long-range propagation, with some detections falling within or near the nominal tropospheric celerity window. We posit two non-exclusive mechanisms to explain this behavior. One involves exceptional propagation conditions, in which acoustic energy is transmitted downwind through highly efficient ducts such as the AtmoSOFAR channel \citep{Albert2023}, allowing rapid, low-loss lateral transmission. The other involves source evolution, whereby significant acoustic energy is generated earlier along the trajectory during phases of intense ablation or distributed shock formation, so that the effective acoustic source precedes or is spatially offset from the point of maximum optical emission. Although unusually efficient downwind propagation through a fast atmospheric duct is our preferred interpretation, either mechanism, individually or in combination, may contribute to the anomalously fast apparent travel times.

In several cases, individual IMS stations record multiple distinct infrasonic arrivals from the same bolide event, implying that acoustic energy can be generated at multiple points along the trajectory rather than originating from a single terminal burst \citep{Brown2011, Silber2014}. Although this behavior is not ubiquitous, its occurrence indicates that dominant acoustic radiation does not always coincide with peak optical brightness and may reflect earlier or spatially separated phases of energy deposition \citep{Brown2011, Pilger2020}. Fragmentation history may contribute to this behavior, but the observations are also consistent with sustained shock generation along an extended path.

Shallow-entry bolides further exhibit greater azimuthal variability across different IMS stations, suggesting that the effective acoustic source is not well represented by a single point location but by a spatially distributed region whose apparent position varies with viewing geometry and propagation path \citep{Pilger2015, Silber2025b}. In contrast, steep-entry bolides tend to produce arrivals with more consistent back azimuths, reflecting a more vertically compact source region and more coherent coupling into atmospheric waveguides. These observations reinforce the conclusion that for shallow entries, peak optical brightness is a poor proxy for the dominant acoustic source region, and that detectability reflects the combined influence of source evolution and atmospheric propagation.

Overall, our results indicate that source geometry and atmospheric propagation act as complementary, probabilistic controls on infrasound detectability. Atmospheric waveguides provide the transmission pathway, but it is the geometry and spatial distribution of the acoustic source that determine whether that pathway is effectively utilized. Successful detection therefore reflects the coincidence of a favorable source configuration and a propagation environment that remains sufficiently coherent to transport acoustic energy to the receiver.

\subsection{Implications for Global Bolide Monitoring and Planetary Defense}\label{planetary_defense}

The results of this study have direct implications for how infrasound observations are interpreted and utilized in planetary defense and global bolide monitoring. The demonstrated detection rate of 50\% for CNEOS bolides represents a substantial increase relative to earlier global surveys, indicating that the contemporary global infrasound system is considerably more capable of monitoring energetic atmospheric entry events than historical baselines suggested.

The results further suggest that detection efficiency may be improved through elevated infrasound sensing, which can access acoustic energy propagating within stratospheric waveguides that are weakly sampled by surface arrays \citep{Bowman2021, Albert2023}. Observations of the 20 April 2023 SpaceX Starship explosion show that energetic acoustic waves can propagate laterally through the AtmoSOFAR channel, an elevated duct near the tropopause, and be recorded by balloon-borne sensors even when surface signals are weak or ambiguous \citep{Albert2023, Bowman2025}. These observations indicate that surface networks sample only a subset of the acoustic field generated by energetic atmospheric events and that elevated sensing platforms may help mitigate geometry- and propagation-driven detection gaps \citep{Young2018, Krishnamoorthy2020, Silber2023}.

Empirical relationships between bolide energy and infrasound signal characteristics provide a useful basis for interpreting infrasonic observations in terms of deposited energy. The results of this study refine these relationships by demonstrating that, although detected bolides are on average more energetic, substantial overlap exists between detected and non-detected populations across the full energy range, and the dependence of detectability on entry angle persists within comparable energy bins. This indicates that while energy controls the overall strength of acoustic generation, it does not uniquely determine detectability, particularly within the moderate-energy regime that dominates the global bolide population. For extreme-energy events, however, total energetics can dominate detectability, such that even geometrically unfavorable entries may produce infrasound observable at global distances, as demonstrated by events such as the 2013 Chelyabinsk superbolide \citep{Brown2013, Popova2013}. Thus, energy governs the potential for infrasound generation, while geometry and atmospheric propagation control whether that energy is efficiently coupled into long-range detectable signals. Importantly, this persistence of geometric dependence within comparable energy bins confirms that entry angle acts as an independent control on detectability, rather than reflecting an indirect correlation with event energetics. Future work could combine object-type classifications with the detection catalog presented here to assess whether asteroidal and cometary populations exhibit different infrasound detection efficiencies.

These geometry-dependent detectability trends have important implications for how infrasound observations are interpreted in hazard and monitoring contexts. Because steep-entry and lower-altitude events are preferentially detected, global infrasound observations provide an incomplete sampling of the full bolide population, particularly for shallow, high-altitude entries that may still carry substantial kinetic energy. As a result, yield estimates and impact flux assessments based solely on infrasound detections may underrepresent the shallow-entry population if geometry-dependent detectability is not considered explicitly. Accounting for these effects is therefore essential when using infrasound-derived statistics to inform assessments of atmospheric entry hazards and impactor populations.

The observed complexity of bolide infrasound signals supports this conclusion. Extended wavetrains, multiple arrivals, and increased azimuthal variability, particularly for shallow-entry events, challenge simple point-source assumptions commonly employed in yield estimation and source localization. For many events, the effective acoustic source is better represented as a distributed region of shock generation that evolves along the trajectory, rather than as a single explosive point coincident with peak optical brightness. Inversion approaches that assume a point source at the altitude of maximum luminosity may therefore misestimate both source location and energy yield, especially for shallow trajectories where the dominant acoustic radiation may be spatially offset from the optical flare.

The kinematic and propagation insights derived here for natural bolides are also directly applicable to the monitoring of artificial atmospheric re-entries. Space debris and controlled spacecraft returns typically enter the atmosphere at shallow angles (<$10^\circ$) and high velocities, closely resembling the grazing fireball population identified here as difficult to detect infrasonically \citep{Silber2024b, Silber2026}. As orbital launch and re-entry activity increases, understanding how trajectory geometry modulates long-range detectability becomes increasingly important for space-traffic monitoring and verification applications \citep[e.g.,][]{Bowman2025, Hatty2026}.

Finally, the geometry-dependent detectability identified in this study suggests that grazing and shallow-entry bolides may be more effectively monitored using regional or dense infrasound networks, rather than relying solely on global-scale systems \citep{Scamfer2026}. Shallow entries deposit acoustic energy over extended horizontal paths and at higher altitudes, conditions that reduce efficient coupling into long-range waveguides but can still produce strong near-field and regional infrasonic signals \citep{Silber2014, Pilger2020, Clemente2025, Silber2026}. Recent studies show that regional infrasound arrays, particularly when combined with adaptive processing and multi-phenomenology observations, are well suited to capturing such events \citep{Silber2025a, Wilson2025, Silber2026}.

The physical principles identified in this study are also relevant to bolide and impact phenomena on other atmosphere-bearing planetary bodies \citep{Petculescu2007, Petculescu2016, Bowman2021, Averbuch2023, Garcia2024}. On planets and moons with substantial atmospheres, such as Venus, Mars, and Titan, hypersonic atmospheric entry and impact events are likewise expected to generate shock-driven acoustic waves whose detectability depends on source geometry, energy deposition altitude, and atmospheric waveguide structure. Although the atmospheric compositions, scale heights, and wind regimes on these bodies differ markedly from those of Earth, the same interplay between entry geometry and propagation environment should govern whether acoustic energy can be transmitted over long distances \citep{Petculescu2007}. In this context, infrasound offers a promising tool for remote sensing of impact events and atmospheric entry processes in extraterrestrial environments, particularly where optical observations are limited \citep[e.g.,][]{Garcia2022}. Future missions equipped with suitable acoustic or pressure sensors could leverage these principles to study impact fluxes, atmospheric structure, and energy deposition processes beyond Earth \citep{Krishnamoorthy2023}. These findings emphasize the complementary roles of global, regional, and elevated infrasound networks in achieving a more complete and less biased characterization of the bolide population, with direct relevance for planetary defense, hazard assessment, and atmospheric entry studies.

A related limitation of this study is that we do not define a single global detectability floor in terms of bolide energy or signal-to-noise ratio. Such a threshold is unlikely to be universal because detection depends jointly on source energy, source altitude and geometry, station distance, array noise conditions, waveform frequency content, and the presence of atmospheric ducts along the source-receiver path. Consequently, lower-energy events may be detected when propagation and station conditions are favorable, whereas more energetic events may remain undetected when acoustic coupling or propagation is unfavorable. A rigorous detectability floor would therefore require a station-level sensitivity model that combines event energy, range, atmospheric propagation, array response, and time-dependent noise conditions. Developing such a probabilistic completeness model is an important direction for future work.

\section{Conclusions}\label{conclusions}

This study presents a systematic global assessment of bolide-generated infrasound using two decades of CNEOS events and modern, multi-frequency array processing applied to the IMS network. We find that infrasound detectability is substantially higher than previously inferred, while remaining fundamentally probabilistic and strongly modulated by source geometry and atmospheric propagation, rather than by entry velocity or energy alone, with energy acting as a secondary modulating factor and source geometry providing the primary control across the dominant moderate-energy regime. In particular, bolides with steep entry angles and lower-altitude energy deposition are preferentially detected, reflecting more favorable coupling of shock-generated acoustic energy into atmospheric waveguides, whereas shallow or grazing entries are less consistently observed at global scales.

The occurrence of fast-arriving phases, multiple arrivals, and azimuthal variability further indicates that bolide infrasound commonly reflects distributed and evolving source regions, rather than a single point coincident with peak optical brightness. While very energetic events can produce detectable infrasound regardless of geometry, for the more common lower-energy regime detectability depends on specific combinations of entry parameters and propagation conditions. These findings support geometry- and propagation-aware interpretations of infrasound energy relations, clarify inherent sampling limitations in global observations, and emphasize the complementary roles of global, regional, and potentially elevated infrasound sensing for planetary defense, atmospheric entry studies, and comparative investigations on atmosphere-bearing planetary bodies. Across the global bolide population, entry angle emerges as the primary geometric control on infrasound detectability, with altitude exerting a secondary influence and entry velocity playing a comparatively minor role.

\section*{CRediT Author Statement}
\textbf{M.\ Ronac Giannone:} Conceptualization, Investigation, Formal Analysis, Visualization, Data Curation, Writing -- Original Draft.
\textbf{E.\ A.\ Silber:} Funding Acquisition, Resources, Project Management, Conceptualization, Formal Analysis, Writing -- Original Draft, Writing -- Review and Editing.

\section*{Data Availability}
The CNEOS fireball data used in this study are publicly available from the NASA Jet Propulsion Laboratory Center for Near-Earth Object Studies at cneos.jpl.nasa.gov/fireballs/. The data that support the findings of this study are available from the CTBTO Preparatory Commission. Restrictions apply to the availability of these data, which were used under license for this study. Data are available from www.ctbto.org/specials/vdec with the permission of the CTBTO Preparatory Commission. The Cardinal detection and processing framework developed and used for this analysis is open-access software and is freely available for download at https://github.com/sjarrowsmith/cardinal.

\section*{Acknowledgements}
The authors thank Ellie Sansom and an anonymous reviewer for their comments that helped improve the paper. We thank Vedant Sawal for assisting with validation of the bolide database through independent identification of infrasound detections. Sandia National Laboratories is a multi-mission laboratory managed and operated by National Technology \& Engineering Solutions of Sandia, LLC (NTESS), a wholly owned subsidiary of Honeywell International Inc., for the U.S. Department of Energy's National Nuclear Security Administration (DOE/NNSA) under contract DE-NA0003525. This written work is authored by an employee of NTESS. The employee, not NTESS, owns the right, title and interest in and to the written work and is responsible for its contents. Any subjective views or opinions that might be expressed in the written work do not necessarily represent the views of the U.S. Government. The publisher acknowledges that the U.S. Government retains a non-exclusive, paid-up, irrevocable, world-wide license to publish or reproduce the published form of this written work or allow others to do so, for U.S. Government purposes. The DOE will provide public access to results of federally sponsored research in accordance with the DOE Public Access Plan.

\section*{Funding}
This work was supported by the Laboratory Directed Research and Development (LDRD) program (project number 229346) at Sandia National Laboratories, a multimission laboratory managed and operated by National Technology and Engineering Solutions of Sandia, LLC., a wholly owned subsidiary of Honeywell International, Inc., for the U.S. Department of Energy's National Nuclear Security Administration under contract DE-NA0003525.

\appendix
\renewcommand{\thefigure}{A\arabic{figure}}
\renewcommand{\thetable}{A\arabic{table}}
\setcounter{figure}{0}
\setcounter{table}{0}

\section{Appendix}\label{appendix}

Cardinal frequency band tables include the following parameters: band (frequency band index), fmin (minimum frequency), fcenter (central frequency), fmax (maximum frequency), win (window length), step (window step). Table A1 shows the values that were used to process each source-receiver pair.

\begin{table}[htbp]
\centering
\caption{Third-octave frequency bands used to process bolide-generated infrasound detected within the global infrasound network.}
\label{tab:A1}
{\footnotesize
\begin{tabular}{@{}llllll@{}}
\toprule
band & fmin (Hz) & fcenter (Hz) & fmax (Hz) & win (s) & step (s) \\ \midrule
1  & 0.02     & 0.022599 & 0.025198 & 164.9825 & 16.49825 \\
2  & 0.025198 & 0.028473 & 0.031748 & 137.0784 & 13.70784 \\
3  & 0.031748 & 0.035874 & 0.04     & 114.9308 & 11.49308 \\
4  & 0.04     & 0.045198 & 0.050397 & 97.35233 & 9.735233 \\
5  & 0.050397 & 0.056946 & 0.063496 & 83.40025 & 8.340025 \\
6  & 0.063496 & 0.071748 & 0.08     & 72.32648 & 7.232648 \\
7  & 0.08     & 0.090397 & 0.100794 & 63.53722 & 6.353722 \\
8  & 0.100794 & 0.113893 & 0.126992 & 56.56118 & 5.656118 \\
9  & 0.126992 & 0.143496 & 0.16     & 51.02429 & 5.102429 \\
10 & 0.16     & 0.180794 & 0.201587 & 46.62966 & 4.662966 \\
11 & 0.201587 & 0.227786 & 0.253984 & 43.14165 & 4.314165 \\
12 & 0.253984 & 0.286992 & 0.32     & 40.3732  & 4.03732  \\
13 & 0.32     & 0.361587 & 0.403175 & 38.17589 & 3.817589 \\
14 & 0.403175 & 0.455572 & 0.507968 & 36.43188 & 3.643188 \\
15 & 0.507968 & 0.573984 & 0.64     & 35.04766 & 3.504766 \\
16 & 0.64     & 0.723175 & 0.806349 & 33.949   & 3.3949   \\
17 & 0.806349 & 0.911143 & 1.015937 & 33.07699 & 3.307699 \\
18 & 1.015937 & 1.147968 & 1.28     & 32.38488 & 3.238488 \\
19 & 1.28     & 1.446349 & 1.612699 & 31.83556 & 3.183556 \\
20 & 1.612699 & 1.822286 & 2.031873 & 31.39955 & 3.139955 \\
21 & 2.031873 & 2.295937 & 2.56     & 31.0535  & 3.10535  \\
22 & 2.56     & 2.892699 & 3.225398 & 30.77883 & 3.077883 \\
23 & 3.225398 & 3.644572 & 4.063747 & 30.56083 & 3.056083 \\
24 & 4.063747 & 4.591873 & 5.12     & 30.3878  & 3.03878  \\
25 & 5.12     & 5.785398 & 6.450796 & 30.25047 & 3.025047 \\
26 & 6.450796 & 7.289145 & 8.127493 & 30.14147 & 3.014147 \\
27 & 8.127493 & 9.183747 & 10       & 30.05496 & 3.005496 \\ \bottomrule
\end{tabular}
}
\end{table}

\begin{figure}[htpb]
\centering
\includegraphics[width=\textwidth]{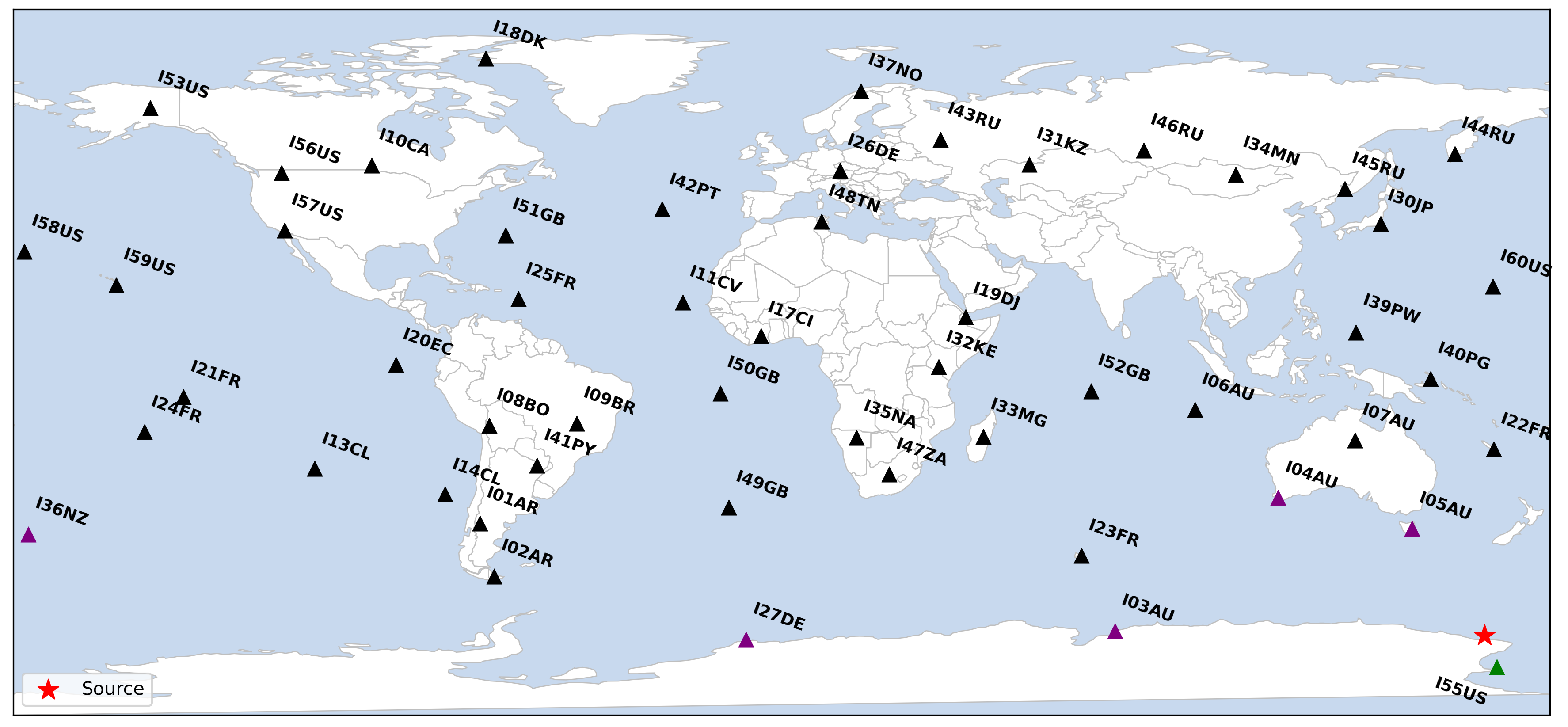}
\caption{Global infrasound network map and the event that occurred on 26 June 2022 at 20:16:26 UTC that was detected by station I55US (bottom right). Green triangles denote stations that were within 1,000 km of the event, while purple triangles represent stations that were within 5,000 km, and black triangles denote stations that exceed 5,000 km from the event.}
\label{fig:A1}
\end{figure}

\begin{figure}[htpb]
\centering
\includegraphics[width=\textwidth]{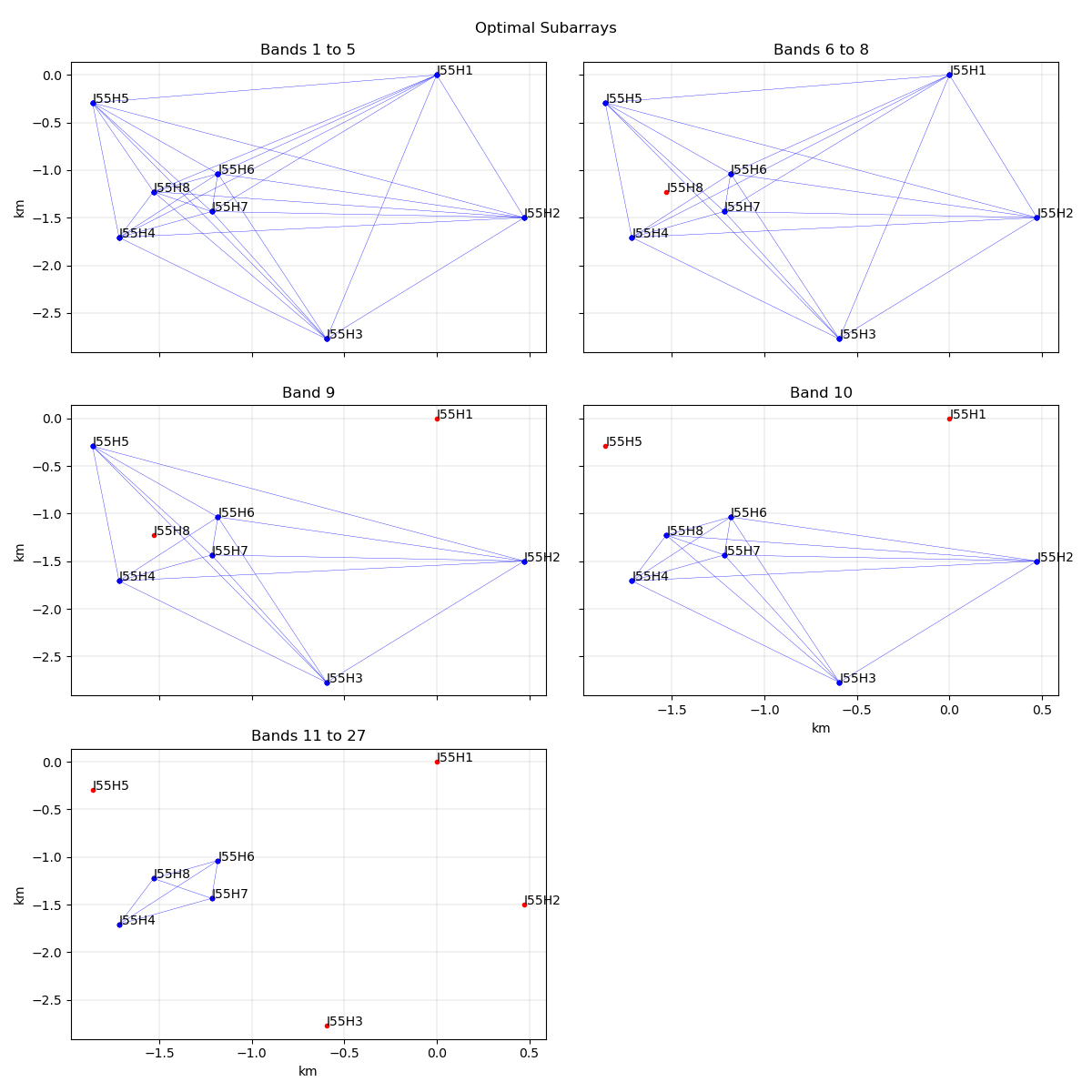}
\caption{Optimal subarray configurations selected for each third-octave frequency band at station I55US. Each panel shows the subset of array elements (blue markers) with connecting lines illustrating the resulting subarray aperture. Sensors excluded from the optimal configuration for that band are shown in red. Lower-frequency bands (top row) favor larger-aperture subarrays, while higher-frequency bands (bottom row) select more compact sensor groupings due to increased sensitivity to coherence loss at higher frequencies.}
\label{fig:A2}
\end{figure}

\begin{figure}[htpb]
\centering
\includegraphics[width=\textwidth]{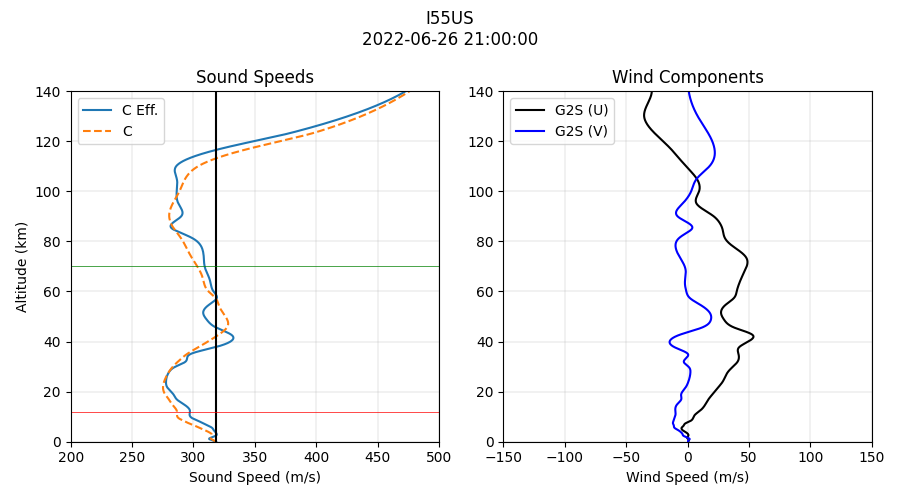}
\caption{Atmospheric profiles at the midpoint between the bolide and station I55US at 21:00 UTC on 26 June 2022. The left panel shows the effective sound speed (solid blue) and adiabatic sound speed (dashed orange), with horizontal lines marking the upper cutoff boundaries of the troposphere (red) and stratosphere (green). The vertical solid black line denotes the effective sound speed at the surface. The right panel shows the corresponding zonal (U) and meridional (V) wind components from the G2S model.}
\label{fig:A3}
\end{figure}

\begin{figure}[htpb]
\centering
\includegraphics[width=\textwidth]{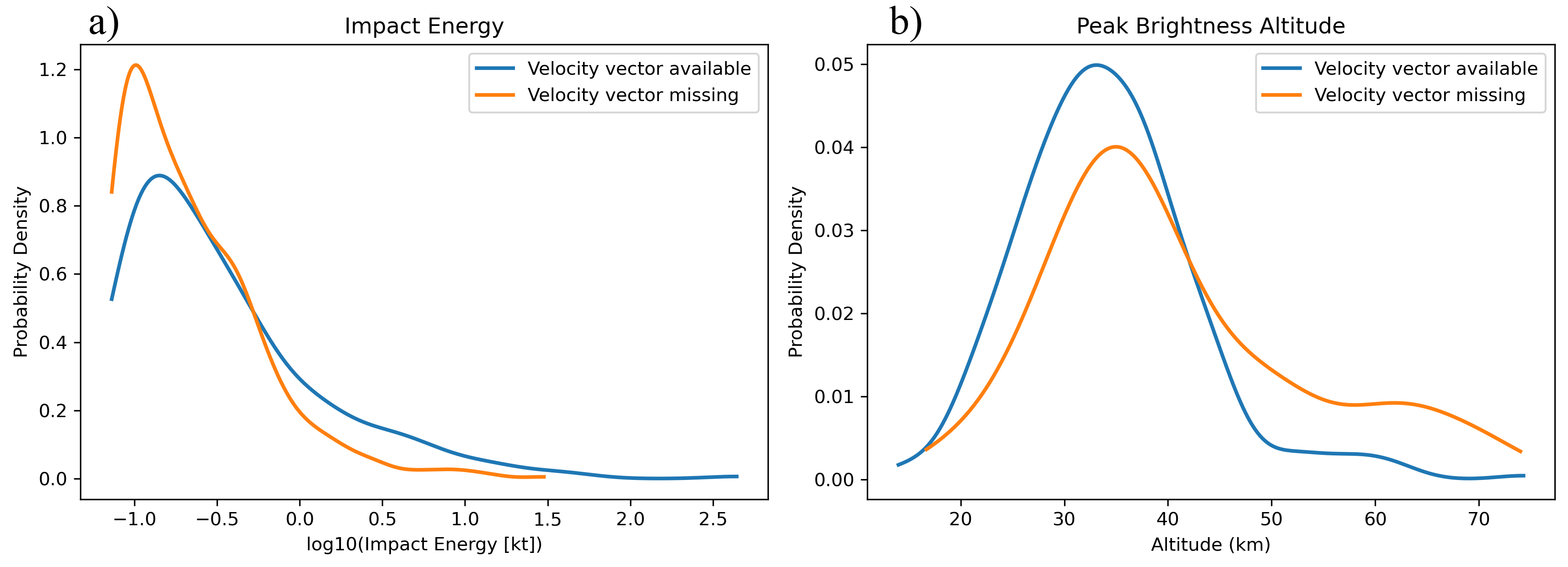}
\caption{Representativeness of the velocity-resolved subset of the CNEOS database. (A) Distribution of impact energy and (B) peak brightness altitude for events with and without available velocity vectors. The velocity-resolved subset is modestly biased toward higher-energy and slightly lower-altitude events relative to the full population.}
\label{fig:A4}
\end{figure}

\begin{figure}[htpb]
\centering
\includegraphics[width=\textwidth]{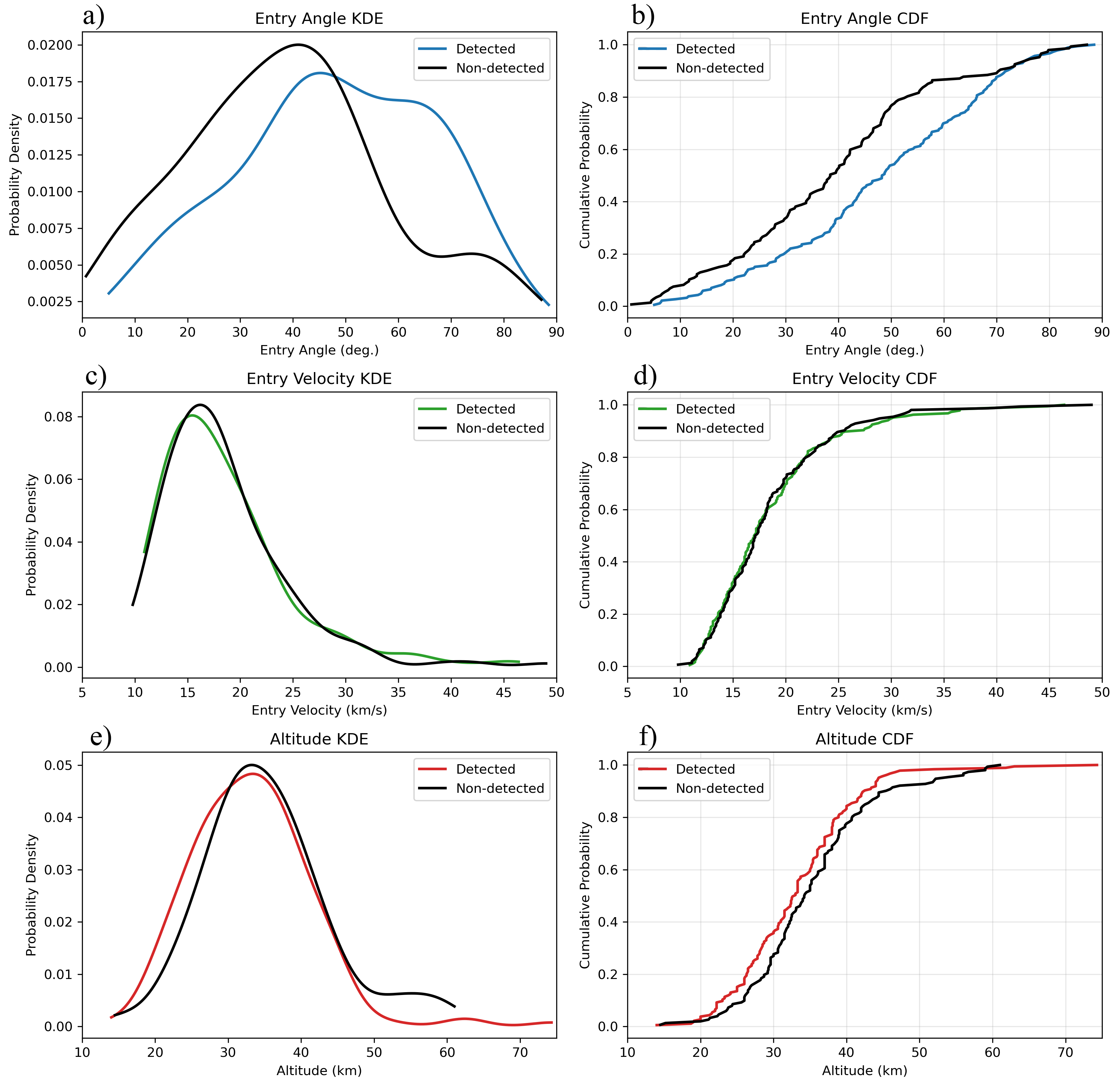}
\caption{Comparison of flight parameters for bolides with available velocity vectors, separated into infrasound-detected and non-detected events. (a,b) Kernel density estimates (KDE) and cumulative distribution functions (CDF) of entry angle show a systematic shift toward steeper angles for detected events. (c,d) Entry velocity distributions are nearly identical between detected and non-detected populations, indicating that velocity alone does not explain detection differences. (e,f) Peak brightness altitude distributions show a modest shift toward lower altitudes for detected events. These results demonstrate that the observed dependence of infrasound detectability on entry angle persists within the velocity-resolved subset and is not an artifact of velocity-vector availability.
}
\label{fig:A5}
\end{figure}

\begin{figure}[htpb]
\centering
\includegraphics[width=\textwidth]{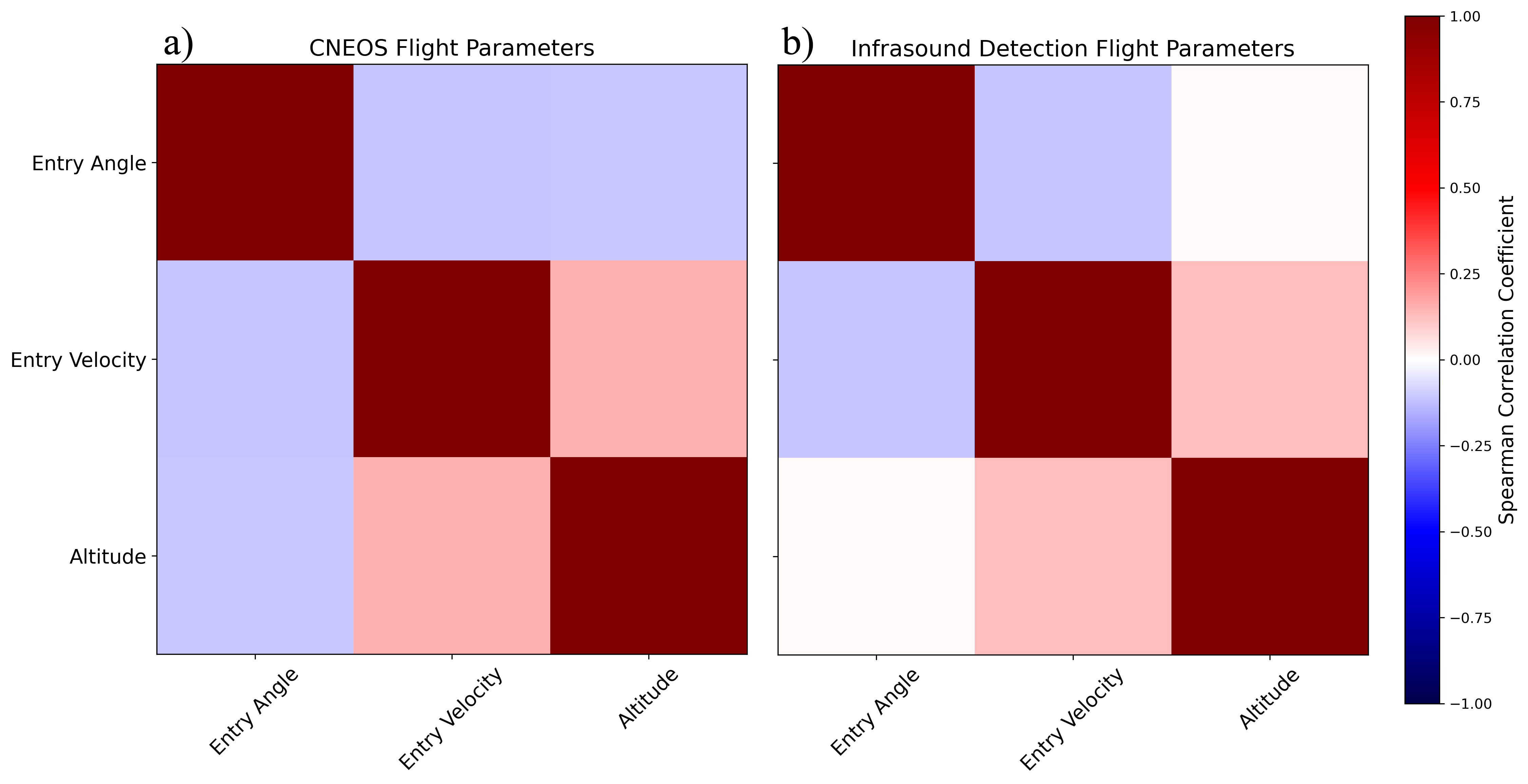}
\caption{Spearman correlation matrices for bolide entry parameters. Panel (a) shows correlations among entry angle, entry velocity, and peak brightness altitude for all CNEOS-reported bolides. Panel (b) plots the correlations for the subset of events that produced infrasound detections. The color scale indicates Spearman correlation coefficients from -1 to 1.}
\label{fig:A6}
\end{figure}

\begin{figure}[htpb]
\centering
\includegraphics[width=\textwidth]{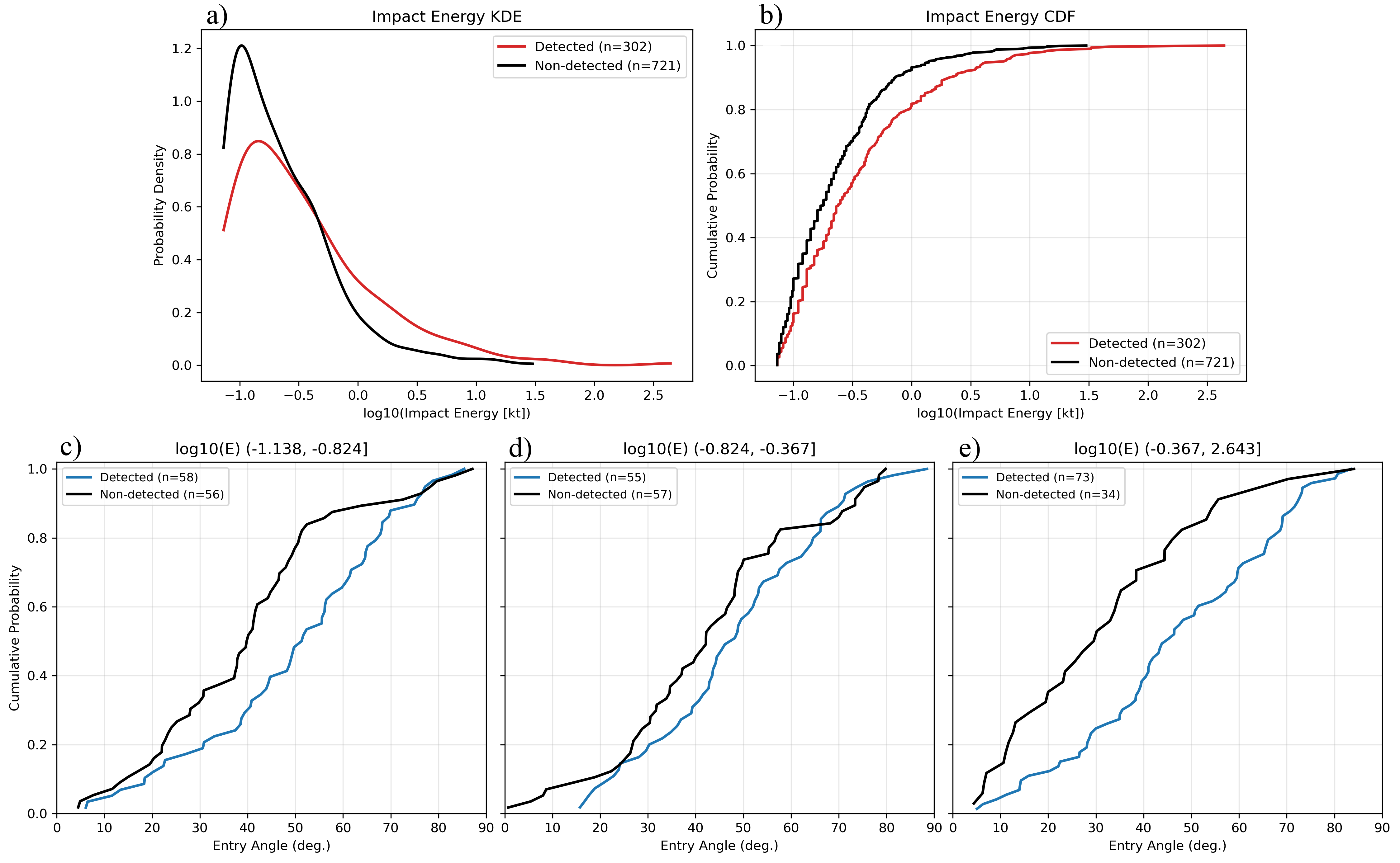}
\caption{Influence of bolide energy on infrasound detectability. (a,b) Impact-energy KDE and CDF comparisons show that detected events are shifted toward higher energies relative to non-detected events, but with substantial overlap between populations. (c--e) Entry-angle CDFs within comparable log$_{10}$(energy) ranges show that detected bolides remain preferentially shifted toward steeper entry angles within each energy subset. These results indicate that energy influences the probability of detection, but the observed entry-angle dependence is not solely an artifact of energy bias.
}\label{fig:A7}
\end{figure}

\begin{figure}[htpb]
\centering
\includegraphics[width=\textwidth]{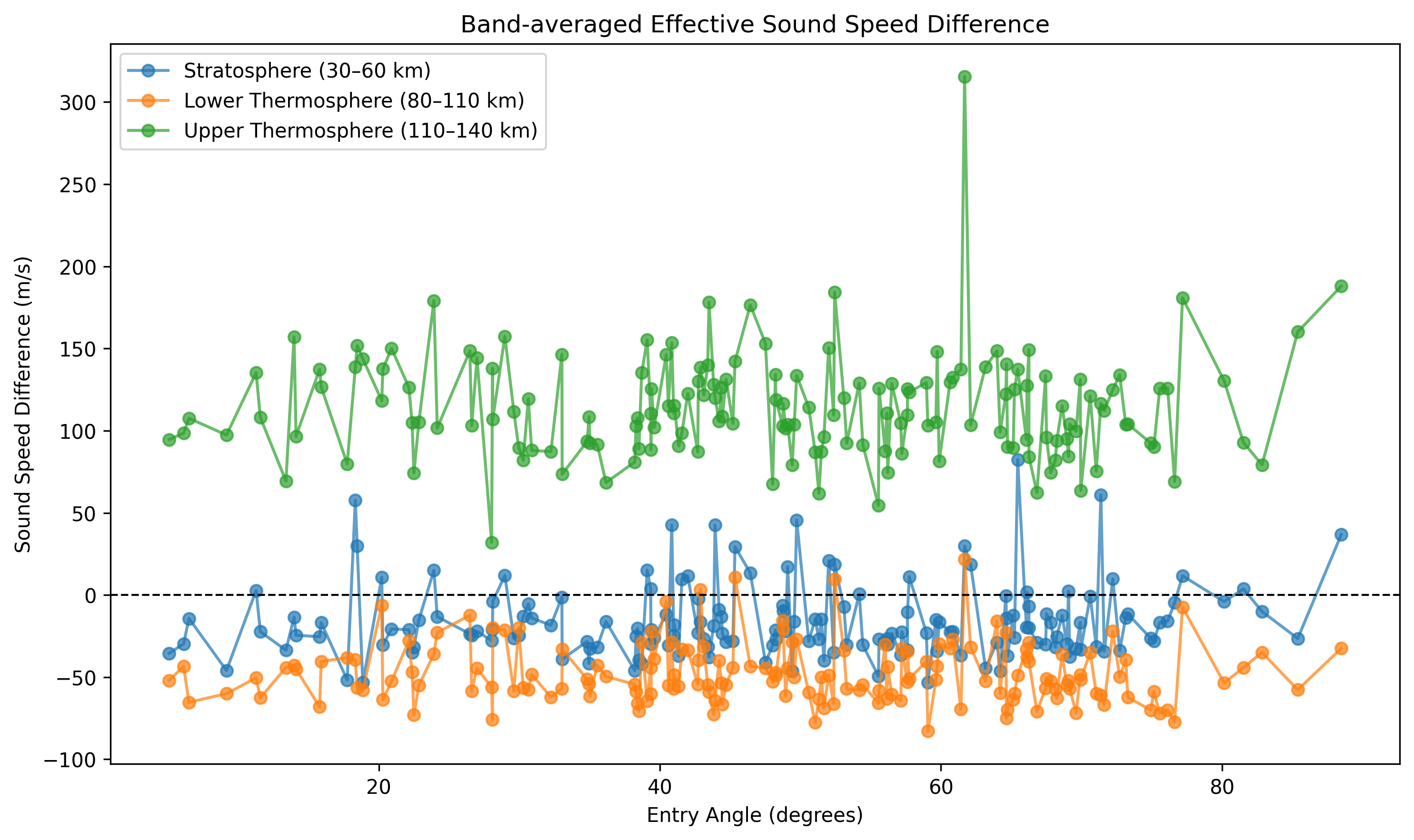}
\caption{Band-averaged effective sound-speed differences as a function of bolide entry angle for all detected events across three atmospheric ducting regimes: the stratosphere (30--60 km), lower thermosphere (80--110 km), and upper thermosphere (110--140 km). Positive values indicate conditions favorable for acoustic refraction back toward the surface, while negative values indicate weak or absent ducting. No systematic relationship between entry angle and duct strength is evident across any altitude band, consistent with the negligible Pearson correlations reported in the text. The substantial scatter highlights the limitations of using range-independent midpoint profiles, which do not capture horizontal variability along the true propagation path.}
\label{fig:A8}
\end{figure}

\end{document}